\documentclass[twocolumn]{aastex631}
\pdfoutput=1

\usepackage{natbib}
\usepackage{graphicx}
\usepackage{hyperref}
\usepackage{graphicx}
\usepackage{float}

\usepackage{siunitx}
\usepackage{amsmath}
\usepackage{appendix}
\usepackage{placeins}
\usepackage{comment}
\usepackage{bm}

\DeclareSIUnit \parsec {pc}
\DeclareSIUnit \year {yr}

\shorttitle{Gaseous Distribution Morphology}
\shortauthors{Qutob et al.}

\begin{document}

\title{Observational Signatures of AGN Feedback in the Morphology and the Ionization States of Milky Way-like Galaxies}

\correspondingauthor{Nadia \ Qutob}
\email{nadiaqutob@gatech.edu}

\author[0000-0001-9500-2639]{Nadia \ Qutob}
\affiliation{Center for Relativistic Astrophysics, School of Physics, Georgia Institute of Technology, 837 State Street, Atlanta, GA 30332-0430, USA}

\author[0000-0002-2791-5011]{Razieh \ Emami}
\affiliation{Center for Astrophysics $\vert$ Harvard \& Smithsonian, 60 Garden Street, Cambridge, MA 02138, USA}

\author[0000-0003-1598-0083]{Kung-Yi \ Su}
\affiliation{Black Hole Initiative at Harvard University, 20 Garden St., Cambridge MA, 02138, USA}

\author[0000-0003-4284-4167]{Randall \ Smith}
\affiliation{Center for Astrophysics $\vert$ Harvard \& Smithsonian, 60 Garden Street,  Cambridge, MA 02138, USA}

\author[0000-0001-6950-1629]{Lars \ Hernquist}
\affiliation{Center for Astrophysics $\vert$ Harvard \& Smithsonian, 60 Garden Street,  Cambridge, MA 02138, USA}

\author[0000-0002-4752-128X]{Dian \ P. \ Triani}
\affiliation{Center for Astrophysics $\vert$ Harvard \& Smithsonian, 60 Garden Street,  Cambridge, MA 02138, USA}

\author[0000-0002-3817-8133]{Cameron Hummels}
\affiliation{TAPIR 350-17, California Institute of Technology, 1200 E. California Boulevard, Pasadena, CA 91125, USA}

\author[0000-0003-3806-8548]{Drummond Fielding}
\affiliation{Center for Computational Astrophysics, Flatiron Institute, 162 Fifth Avenue, New York, NY 10010, USA}

\author[0000-0003-3729-1684]{Philip F. Hopkins}
\affiliation{TAPIR 350-17, California Institute of Technology, 1200 E. California Boulevard, Pasadena, CA 91125, USA}

\author{Rachel S. Somerville}
\affiliation{Center for Computational Astrophysics, Flatiron Institute, 162 Fifth Avenue, New York, NY 10010, USA}

\author[0000-0001-8128-6976]{David R. Ballantyne}
\affiliation{Center for Relativistic Astrophysics, School of Physics, Georgia Institute of Technology, 837 State Street, Atlanta, GA 30332-0430, USA}

\author[0000-0001-8593-7692]{Mark \ Vogelsberger}
\affiliation{Department of Physics, Kavli Institute for Astrophysics and Space Research, Massachusetts Institute of Technology, Cambridge, MA 02139, USA}

\author[0000-0002-5445-5401]{Grant \ Tremblay}
\affiliation{Center for Astrophysics $\vert$ Harvard \& Smithsonian, 60 Garden Street, Cambridge, MA 02138, USA}

\author[0000-0002-5872-6061]{James F. Steiner}
\affiliation{Center for Astrophysics $\vert$ Harvard \& Smithsonian, 60 Garden Street, Cambridge, MA 02138, USA}

\author[0000-0003-2808-275X]{Douglas Finkbeiner}
\affiliation{Center for Astrophysics $\vert$ Harvard \& Smithsonian, 60 Garden Street, Cambridge, MA 02138, USA}

\author{Ramesh Narayan}
\affiliation{Center for Astrophysics $\vert$ Harvard \& Smithsonian, 60 Garden Street, Cambridge, MA 02138, USA}
\affiliation{Black Hole Initiative at Harvard University, 20 Garden St., Cambridge MA, 02138, USA}

\author{Minjung Park}
\affiliation{Center for Astrophysics $\vert$ Harvard \& Smithsonian, 60 Garden Street, Cambridge, MA 02138, USA}

\author{Josh Grindlay}
\affiliation{Center for Astrophysics $\vert$ Harvard \& Smithsonian, 60 Garden Street, Cambridge, MA 02138, USA}

\author[0000-0002-5554-8896]{Priyamvada Natarajan}
\affiliation{Black Hole Initiative at Harvard University, 20 Garden St., Cambridge MA, 02138, USA}
\affiliation{Department of Astronomy, Yale University, New Haven, CT 06511, USA}
\affiliation{Department of Physics, Yale University, New Haven, CT 06520, USA}

\author[0000-0003-4073-3236]{Christopher C. Hayward}
\affiliation{Center for Computational Astrophysics, Flatiron Institute, 162 Fifth Avenue, New York, NY 10010, USA}

\author{Du\v{s}an Kere\v{s}}
\affiliation{Department of Physics \& Astronomy and CIERA, Northwestern University, 1800 Sherman Ave, Evanston, IL 60201, USA}

\author[0000-0002-7484-2695]{Sam B. Ponnada}
\affiliation{TAPIR 350-17, California Institute of Technology, 1200 E. California Boulevard, Pasadena, CA 91125, USA}

\author{Sirio Belli}
\affiliation{Dipartimento di Fisica e Astronomia, Università di Bologna, Bologna, Italy}

\author{Rebecca Davies}
\affiliation{Centre for Astrophysics and Supercomputing, Swinburne University of Technology, Hawthorn, Victoria, Australia} 
\affiliation{ARC Centre of Excellence for All Sky Astrophysics in 3 Dimensions (ASTRO 3D), Australia}

\author{Gabriel Maheson}
\affiliation{Kavli Institute for Cosmology, University of Cambridge, Madingley Road, Cambridge CB3 0HA, UK}
\affiliation{Cavendish Laboratory, University of Cambridge, 19 J. J. Thomson Ave., Cambridge CB3 0HE, UK}

\author{Letizia Bugiani}
\affiliation{Dipartimento di Fisica e Astronomia, Università di Bologna, Bologna, Italy}

\author[0000-0002-0682-3310]{Yijia Li}
\affiliation{Department of Astronomy \& Astrophysics, The Pennsylvania State University, University Park, PA 16802, USA}
\affiliation{Institute for Gravitation and the Cosmos, The Pennsylvania state University, University Park, PA 16802, USA}

\begin{abstract}

\noindent 
We make an in-depth analysis of different AGN jet models' signatures, inducing quiescence in galaxies with a halo mass of $10^{12} M_\odot$. Three jet models, including cosmic ray-dominant, hot thermal, and precessing kinetic jets, are studied at two energy flux levels each, compared to a jet-free, stellar feedback-only simulation. We examine the distribution of Mg {\small II}, O {\small VI}, and O {\small VIII} ions, alongside gas temperature and density profiles. 
Low-energy ions, like Mg {\small II}, concentrate in the ISM, while higher energy ions, e.g., O {\small VIII}, prevail at the AGN jet cocoon's edge. High-energy flux jets display an isotropic ion distribution with lower overall density. High-energy thermal or cosmic ray jets pressurize at smaller radii, significantly suppressing core density. The cosmic ray jet provides extra pressure support, extending cool and warm gas distribution. A break in the ion-to-mass ratio slope in O {\small VI} and O {\small VIII} is demonstrated in the ISM-to-CGM transition (between 10-30 kpc), growing smoothly towards the CGM at greater distances.

\end{abstract}


\section{Introduction} \label{sec:intro}
Massive elliptical galaxies and galaxy clusters are distinctly characterized by their red and dormant state, indicative of an overarching deficiency in star formation. Recent X-ray observations  \citep{1994ApJ...436L..63F,2019MNRAS.488.2549S} affirm that the radiative cooling of hot gas within galaxy clusters should in principle 
induce the cooling and condensation of circumgalactic medium (CGM) gas into the interstellar medium (ISM), fostering star formation at a significantly elevated rate compared to current observations. The persistent incongruity between the expected dynamics of gas cooling and the observed scarcity of star formation in massive galaxies, particularly those with halo masses exceeding $10^{12} M_{\odot}$, is formally identified as the ``cooling flow problem." 

One potential avenue to unraveling this conundrum involves the strategic infusion of substantial energies from the inner reaches of the galaxy via Active Galactic Nuclei (AGN) feedback. This deliberate energy injection into both the ISM and CGM is envisioned as a means to forestall gas cooling processes. Recent works have elucidated scenarios wherein the injected energy, particularly through AGN feedback in the form of jets \citep{2018MNRAS.473L.111S,2019MNRAS.487.4393S,2021MNRAS.507..175S}, can induce a cessation of star formation, inducing a state of quiescence within the system.
Specifically, the influence of AGN feedback, notably in the manifestation of jets (see also \citealt[][]{2012ApJ...746...94G,2014ApJ...789...54L,2016ApJ...818..181Y,2017MNRAS.472.4707B,2017ApJ...844...13R,2019MNRAS.483.2465M}), or alternative modes of AGN feedback \citealt[][]{2017ApJ...837..149G,2017MNRAS.468..751E,2018MNRAS.479.4056W,2018ApJ...866...70L,2018ApJ...856..115P,2018ApJ...864....6Y,2020NatRP...2...42V}, has emerged as a focal point in these inquiries.

\begin{figure}[th!]
\center
\includegraphics[width=0.6\textwidth]{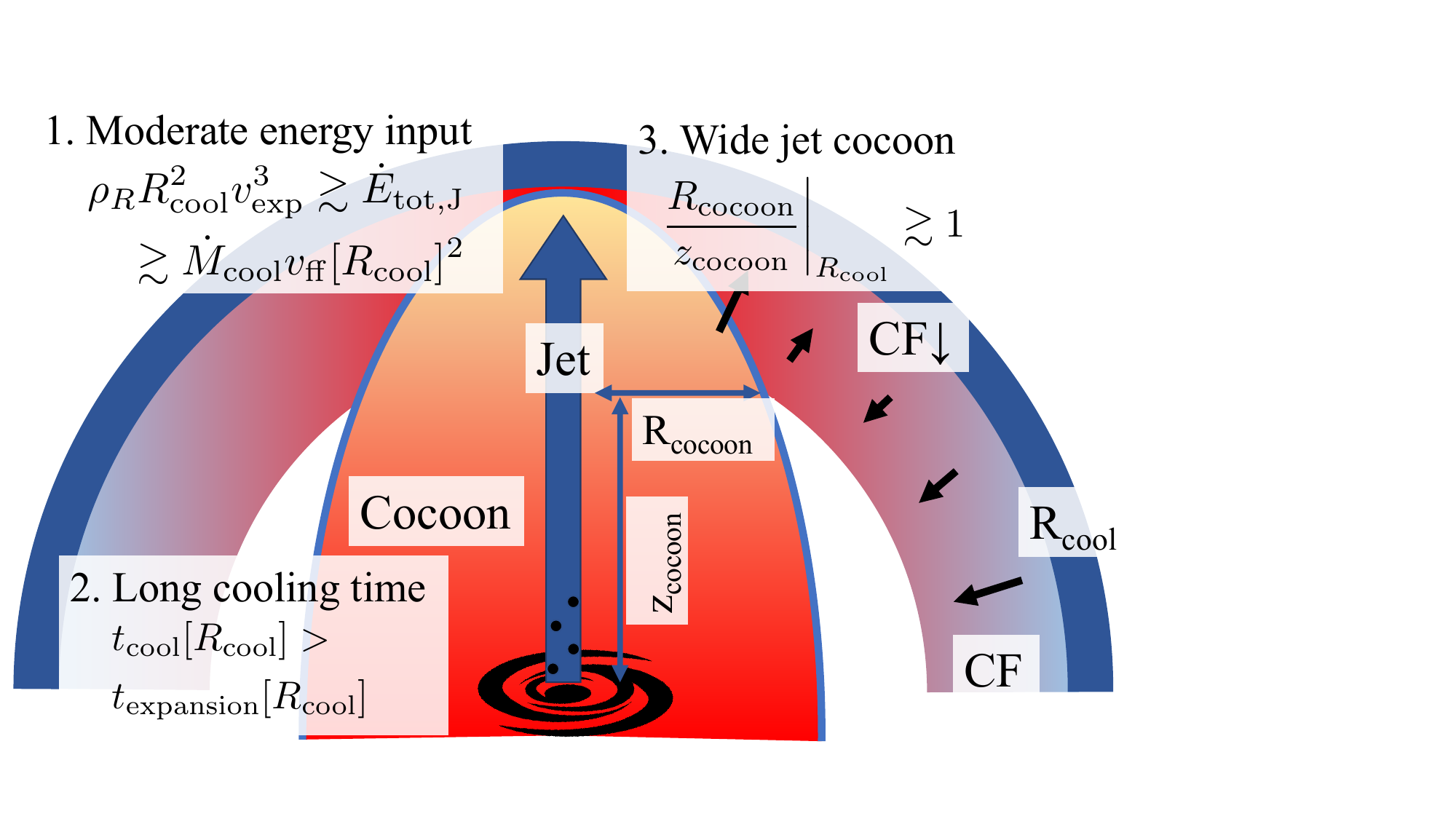}
\caption{A cartoon picture of the criteria for successful jet models. Depicts AGN jets counterbalancing the cooling flow of the CGM. }
\label{fig:AGN-feedback-cartoon}
\end{figure}

Earlier studies proposed specific properties for AGN jets to effectively quench star formation. \cite{2021MNRAS.507..175S, 2023arXiv231017692S} systematically examined various jet models, adjusting parameters such as mass, energy flux, energy composition, and jet precession. The investigation yielded three key criteria for a successful jet model as depicted in Figure \ref{fig:AGN-feedback-cartoon}.
Firstly, the jet should exhibit an energy flux comparable to the free-fall energy flux at the cooling radius of the halo, hindering the cooling and inflow of gas toward the halo center. Insufficient energy flux renders the jet insufficiently powerful for quenching. Secondly, the jet must generate a sufficiently broad cocoon. An overly narrow cocoon results in uneven energy distribution, limiting its efficacy in quenching star formation. Thirdly, effective quenching necessitates a sufficiently extended cooling time for the jet. A brief cooling period prevents the jet from depositing ample energy before dissipating through radiation. The cooling time ($t_{\rm cool}\sim kT/\bar{n}\Lambda(T)$) should notably surpass the cocoon expansion time ($t_{\rm exp}\sim R_{\rm cool}/v_{\rm exp}$) \cite{AGN_Jet_halo_mass}, emphasizing the intricate balance of parameters crucial for AGN jets to quell star formation in galactic environments.

\cite{AGN_Jet_halo_mass} demonstrated that cosmic ray jets, precessing kinetic energy jets, and hot thermal jets, when endowed with sufficient energy fluxes, consistently meet the criteria for quenching star formation. Importantly, these models do not blatantly violate X-ray observations. The extended cooling time in these models, attributed to cosmic ray energy or high specific heat, surpasses the cocoon expansion duration. Pressure generated by thermal or cosmic ray energy, coupled with jet precession, effectively widens the jet cocoon at cooling radii, strategically impeding cooling flows across a broad solid angle. Despite recent advancements in identifying jet categories that fulfill the criteria for successful galaxy quenching, achieving a comprehensive understanding of their specific impacts on various galactic properties remains elusive. This knowledge gap hampers our capacity to effectively leverage observations of multi-phase gas.

This paper extends upon the advancements achieved in the AGN feedback models presented in \cite{2021MNRAS.507..175S} by establishing connections between AGN jet models and CGM properties, including gas density, temperature, and ionization state. Three distinct jet models — cosmic ray jets, precessing kinetic energy jets, and hot thermal jets — will be employed in a subset of isolated Milky Way-like galaxies, characterized by a halo mass of $10^{12}M_\odot$. Our focus in this study will center on the morphologies of three distinct ions including Mg {\small II}, O {\small VI}, and O {\small VIII}, each characterized by significantly different ionization energies. This choice ensures effective sampling of the gas within each distinct ionization state. The overarching objective is to investigate the broader impact of AGN jets on the distribution of gas temperature and metallicity. These factors collectively contribute to modifying the ion distributions within both the CGM and the ISM.

The paper is structured as follows. In Section \ref{sec: method} we explain the method used to make the synthetic observations. Section \ref{sec:results} details the findings for each type of simulation and parameter setting. In section \ref{sec: discussion} we summarize the results and describe future directions. Section \ref{conc} presents the conclusion of this manuscript. 

\section{Methodology} 
\label{sec: method}
Our investigation delves into the observational manifestations of diverse AGN jet models by discerning the ion distribution in a subset of Milky Way-like isolated galaxy simulations. These simulations incorporate AGN jet models, firstly introduced in \citet{2021MNRAS.507..175S, 2023arXiv231017692S}. Post-processing of the simulations is conducted using the synthetic absorption toolkit {\small TRIDENT} \citep{2017ApJ...847...59H}. Subsequently, we explore the morphologies of various ions spanning low-to-high ionization states. The synthesis of simulations and post-processing procedures is summarized as follows.

\subsection{Simulations} 
\label{sec: observations} 
\begin{figure*}[th!]
\centering
\includegraphics[scale=0.38]{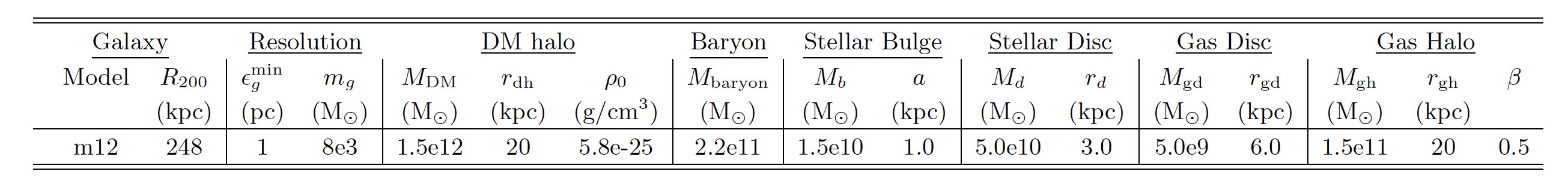}
\caption{Parameters of the galaxy/halo model under investigation in this paper: 
{\color{purple} (1)} Model name. The numerical suffix following `m' designates the approximate logarithmic halo mass.
{\color{purple} (2)} \textbf{$R_{\rm 200}$}: The radius encompassing an average density of 200 times the critical density.
{\color{purple} (3)} \textbf{$\epsilon_g^{\rm min}$}: Minimum gravitational force softening for gas. The adaptive softening for gas in all simulations is matched to the hydrodynamic resolution; presented here is the minimum Plummer equivalent softening.
{\color{purple} (4)} \textbf{$m_g$}: Gas mass (resolution element). For m14, there exists a resolution gradient, and its 
$m_g$ corresponds to the mass of the highest resolution elements.
{\color{purple} (5)} \textbf{$M_{\rm halo}$}: Halo mass. 
{\color{purple} (6)} \textbf{$r_{\rm dh}$}: NFW halo scale radius (the corresponding concentration of m14 is $c=5.5$).
{\color{purple} (7)} \textbf{$V_{\rm max}$}: Halo maximum circular velocity.
{\color{purple} (8)} \textbf{$M_{\rm baryon}$}: Total baryonic mass. 
{\color{purple} (9)} \textbf{$M_b$}: Bulge mass.
{\color{purple} (10)} \textbf{$a$}: Bulge Hernquist-profile scale-length.
{\color{purple} (11)} \textbf{$M_d$}: Stellar disc mass.
{\color{purple} (12)} \textbf{$r_d$}: Stellar disc exponential scale-length.
{\color{purple} (13)} \textbf{$M_{\rm gd}$}: Gas disc mass. 
{\color{purple} (14)} \textbf{$r_{\rm gd}$}: Gas disc exponential scale-length.
{\color{purple} (15)} \textbf{$M_{\rm gh}$}: Hydrostatic gas halo mass. 
{\color{purple} (16)} \textbf{$r_{\rm gh}$}: Hydrostatic gas halo $\beta=1/2$ profile scale-length.
} \label{tab:ic}
\end{figure*}

The isolated simulation library forming the foundation of our analysis employs {\small GIZMO} \citep{2015MNRAS.450...53H} in its MFM (meshless finite mass) mode. This Lagrangian mesh-free Godunov method seamlessly integrates the benefits of both grid-based and smoothed-particle hydrodynamics (SPH) methods. A thorough series of methodological papers, including those on hydrodynamics and self-gravity \citep{2015MNRAS.450...53H}, magnetohydrodynamics \citep[MHD;][]{2016MNRAS.455...51H,2016MNRAS.462..576H}, anisotropic conduction and viscosity \citep{2017MNRAS.466.3387H,2017MNRAS.471..144S}, and cosmic rays \citep{Chan_2019}, provide exhaustive details on extensive tests and numerical implementations. All simulations incorporate the FIRE-2 implementation, featuring comprehensive treatments of the ISM, star formation, and stellar feedback as outlined in \citet{2018MNRAS.477.1578H,2018MNRAS.480..800H}. The cooling mechanism spans a temperature range of \(10^1-10^{10}\)K, encompassing the effects of photo-electric and photo-ionization heating, fine structure, collisional, Compton, recombination, atomic, and molecular cooling. The initial conditions for the isolated galaxy simulations are provided in detail in Table \ref{tab:ic} and are further elucidated upon in Section \ref{Initial-Conditions}.

The treatment of star formation in these simulations employs a sink particle method, restricted to molecular, self-shielding, and locally self-gravitating gas with a density 
$n>100\,{\rm cm^{-3}}$ \citep{2013MNRAS.432.2647H}. Once formed, star particles in the simulation represent a single stellar population, inheriting their metallicity from the parent gas particle at the formation stage. Feedback rates, including supernovae (SNe) and mass-loss rates, are averaged over the initial mass function (IMF) values calculated from {\small STARBURST99} \citep{1999ApJS..123....3L} using a \citet{2002Sci...295...82K} IMF.

The stellar feedback model encompasses three key components: (1) The Radiative feedback, incorporating photo-ionization and photo-electric heating, along with radiation pressure tracked in five bands (ionizing, FUV, NUV, optical-NIR, IR), (2) Continuous stellar mass loss and injection of mass, metals, energy, and momentum from O star, B star, and asymptotic giant branch (AGB) winds, and (3) Type II and Ia Supernovae, occurring based on tabulated rates and injecting the appropriate mass, metals, momentum, and energy into the surrounding gas. Our simulations additionally incorporate magnetohydrodynamics (MHD), fully anisotropic conduction, and viscosity with Spitzer-Braginski coefficients.

\subsection{Initial Conditions} 
\label{Initial-Conditions}
The initial conditions (ICs) scrutinized in this study are extensively presented and elucidated in \cite{2019MNRAS.487.4393S,2021MNRAS.507..175S,2023arXiv231017692S} and are summarized in Tables \ref{tab:ic} and \ref{tab:run}. These ICs are tailored to resemble a Milky Way-like galaxy, assuming the presence of no missing baryons within approximately one virial radius. The initialization of the dark matter (DM) halo, bulge, black hole, and gas+stellar disk adheres to the prescriptions of \cite{1999MNRAS.307..162S} and \cite{2000MNRAS.312..859S}.

We adopt the following components for the ICs: a spherical, isotropic, NFW (Navarro–Frenk–White) \citep{1996ApJ...462..563N} profile DM halo; a \cite{1990ApJ...356..359H} profile stellar bulge; an exponential, rotation-supported disk of gas and stars initialized with Toomre 
$Q\approx1$; a black hole (BH) with a mass approximately $\sim1/300$ of the bulge mass \citep[e.g.,][]{2004ApJ...604L..89H}; and an extended spherical, hydrostatic gas halo with a 
$\beta$-profile and rotation at twice the net DM spin (i.e., $\sim 10-15\%$  of the support against gravity arises from rotation, with the majority of support from thermal pressure as anticipated in a massive halo). Table \ref{tab:ic} summarizes key properties of the initial dark matter halo, stellar bulge, stellar disk, ISM gas disk, and gaseous CGM halo.

The initial metallicity of the CGM/ICM drops from solar ($Z=0.02$) to $Z=0.001$  obeying the functional form: $Z(r)=0.02\,(0.05+0.95/(1+(r/r_Z\,{\rm })^{1.5}))$, where $r_Z=20$ kpc. The boundary between the ISM and CGM is defined at $r_Z=20$ kpc. 
Initial magnetic fields are azimuthal with a seed value of $|{\bf B}|=B_0/(1+(r/r_B)^{2})^{\beta_B}$ extending throughout the CGM, where $B_0=0.03 {\rm \mu G}$, $r_B=20 {\rm kpc}$, and  $\beta_B=0.375$. The initial CR energy density is in equipartition with the local initial magnetic energy density. 

\subsection{Types of AGN Jets} 
\label{sec: AGN-models} 
In each simulation, the initiation of the jet involves a particle spawning method, generating new gas cells (resolution elements) originating from the central black hole. These spawned gas particles exhibit a mass resolution of 5000 ${\rm M}_\odot$ and are explicitly restricted from de-refining (merging into a regular gas element) until their deceleration reaches 10\% of the launch velocity. Once a sufficient level of deceleration is attained, de-refinement becomes permissible. To ensure exact linear momentum conservation, two particles are simultaneously spawned in opposite z-directions when the cumulative jet mass flux reaches twice the targeted spawned particle mass.

\begin{figure*}[th!]
\centering
\includegraphics[scale=0.38]{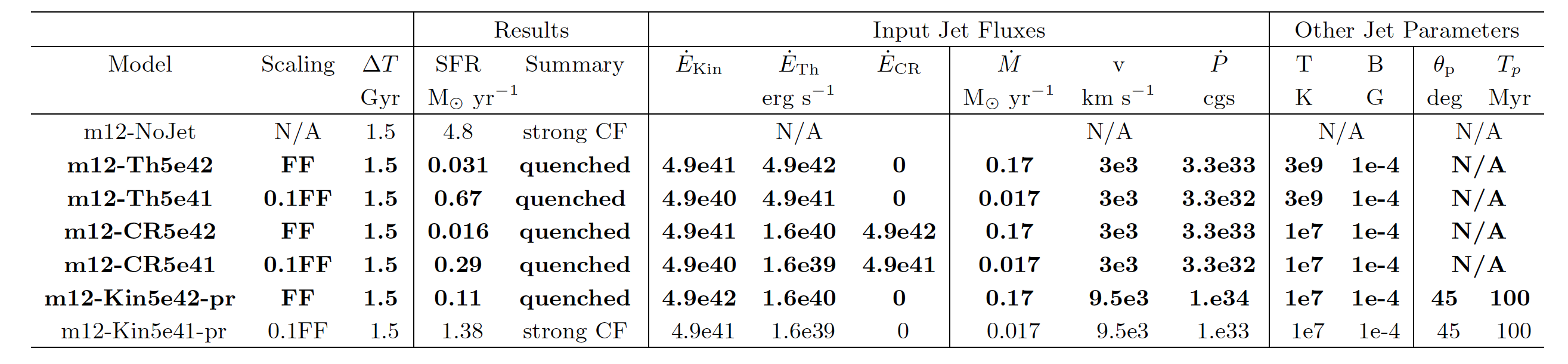}
\caption{ The description of each jet model and the corresponding defining parameters is as follows:\\
{\color{blue} (1)} Model name:  The nomenclature for each model begins with the primary form of energy flux followed by the specific value in erg s$^{-1}$ utilized. Additionally, the label `pr' denotes the inclusion of jet precession in the model designation.
{\color{blue} (2)} Scaling: The scaling of the energy flux was achieved by adjusting the mass flux while maintaining the specific energy in accordance with the free-fall energy within $R_{\rm cool}$ (`FF'). Another iteration was conducted with 0.1 times that energy flux (`0.1FF').
{\color{blue} (3)} $\Delta T$: The duration of the simulations extends to $1.5 \mathrm{Gyr}$, unless either when the halo is entirely ``blown out" or when it remains entirely unaffected.
{\color{blue} (4)}  The SFR averaged over the last \(250 \mathrm{Myr} \).
{\color{blue} (5)}  Summary of the results: `strong CF', and `quenched' correspond respectively to sSFR of $\gtrsim 10^{-11}$,  and $\lesssim 10^{-11} {\rm yr}^{-1}$.
{\color{blue} (6)}  $\dot{E}_{\rm Kin}$, $\dot{E}_{\rm Th}$, and $\dot{E}_{\rm CR}$ tabulate the total energy input of the corresponding form. 
{\color{blue} (7)}  $\dot{M}$, v, and $\dot{P}$ tabulate the mass flux, jet velocity, and momentum flux. 
{\color{blue} (8)}  T refers to the initial temperature of the jet. 
{\color{blue} (9)}  B is the maximum initial magnetic field strength of the jet; (t) and (p) means toroidal and poloidal, respectively. 
{\color{blue} (10)} $\theta_{p}$: The precession angle of the jet. 
{\color{blue} (11)} $T_{p}$: Precession period. 
} \label{tab:run}
\end{figure*}

The jet features a 1$^\circ$ opening angle around the jet axis, aligned along the z-axis, with the exception of the precessing-kinetic jet. In the latter, the jet axis undergoes precession around the z-axis at a $45^\circ$ angle with a periodicity of 100 Myr. Jet particles are generated with a fixed mass flux, and specific values for temperature, velocity, magnetic fields, and cosmic ray energy are assigned based on the chosen jet model. These parameters, in turn, determine the kinetic, thermal, and cosmic energy fluxes, as outlined in Table \ref{tab:run}.

\citet{2021MNRAS.507..175S,2023arXiv231017692S} concluded that, within the halo mass range of $10^{12}-10^{15} M_\odot$, successful jet models must exhibit specific characteristics including a wide jet cocoon and a sufficiently extended cooling time. They identified preferred models among these criteria, favoring either a high-temperature thermal jet, a cosmic ray (CR) dominant jet, or a widely precessing-kinetic jet. Consequently, our focus narrows to these three jet models. Each model is assessed under two distinct energy fluxes: one with high and the other with low energy flux. Simulation runs featuring higher jet energy fluxes are scaled from the energy flux necessary to quench a $10^{14}M_\odot$ halo based on the free fall energy flux at the cooling radius. Conversely, runs with lower fluxes exhibit only 10\% of the energy flux found in their high-flux counterparts. Table \ref{tab:run} provides a comprehensive summary of the fluxes in all simulation runs in various forms.

We conduct a crucial comparison between each of the three identified jets and a no-jet simulation exclusively incorporating stellar feedback, serving as a control. This comparative analysis enables a comprehensive examination of AGN jet behavior and its influence on the surrounding ISM CGM for quenching effects. In the absence of an AGN jet, the stellar feedback alone is insufficient to suppress star formation, as the energy flow does not counterbalance the cooling from the adjacent CGM. Given the central objective of this project to scrutinize the impact of AGN jets on galactic quenching, the juxtaposition of each of our three selected jets, under both high and low energy flux conditions, against an unquenched galaxy with no jet is indispensable. This comparative approach is essential for discerning significant trends in energy and mass distribution, providing insights that can be translated to observables. The outcome comprises a total of seven simulations post-processed for each of the ions under investigation. Further details regarding this comparative analysis are presented in Section \ref{sec:results}.

\begin{figure*}[th!]
\centering
\includegraphics[scale=0.47]{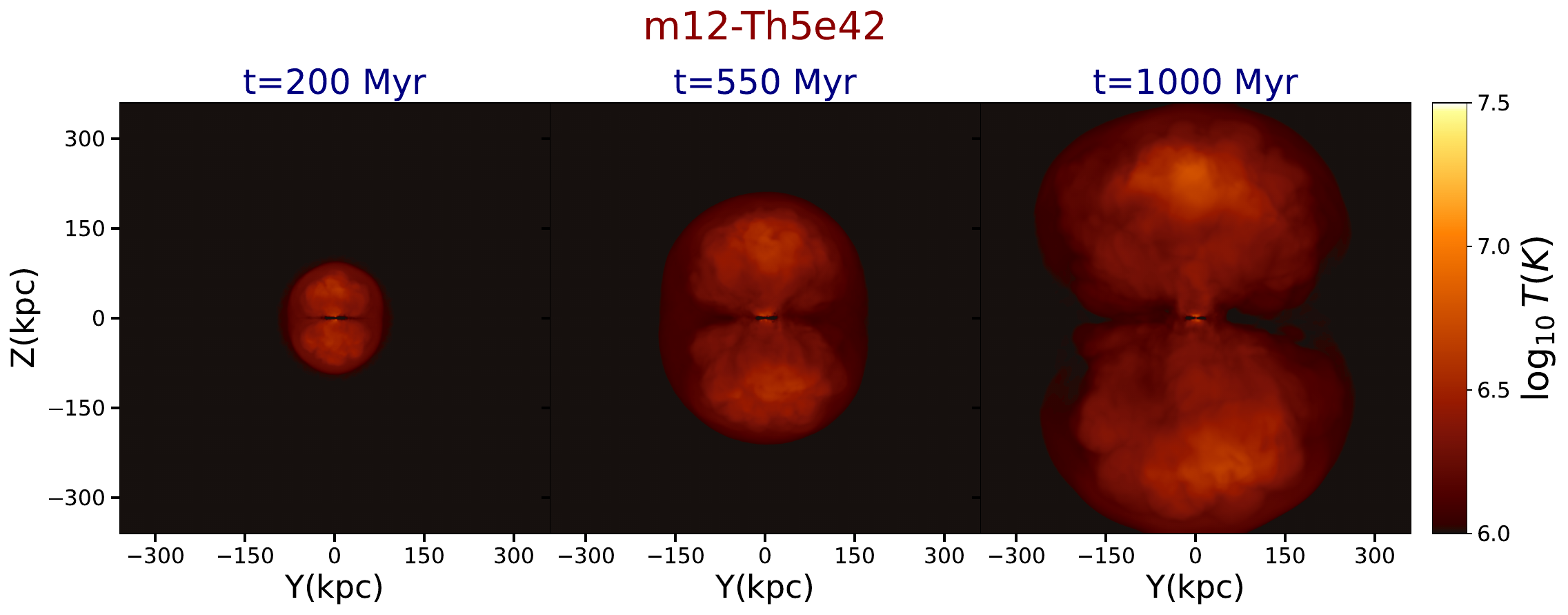}
\caption{The thermal jet propagation at several distinct snapshots. Each panel displays the mass-weighted temperature map in the Y-Z projection.}
\label{fig:Jet propagation}
\end{figure*}    

\begin{figure*}[th!]
\centering
\includegraphics[scale=0.34]{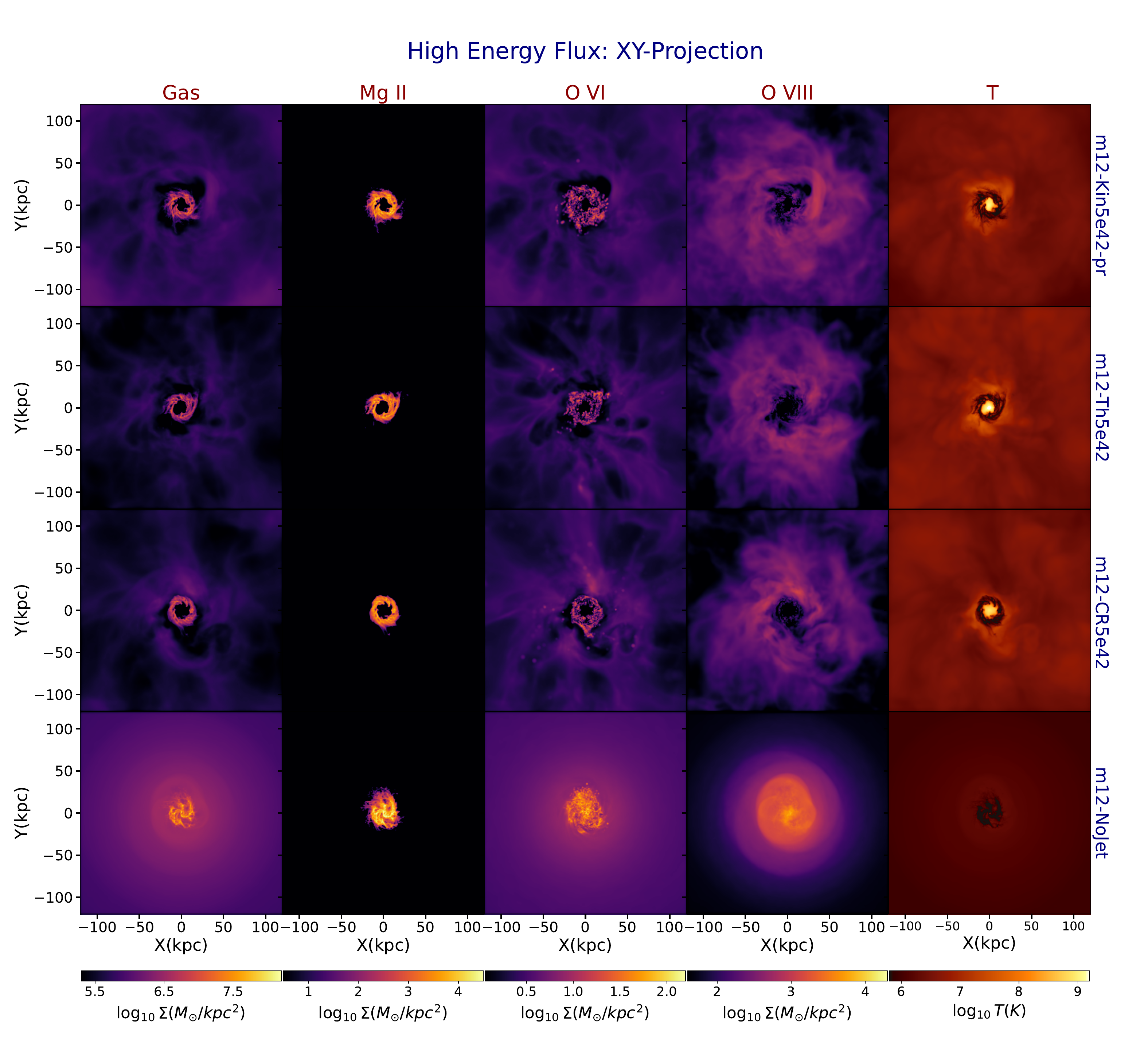}
\caption{The face-on (XY-projected) view of different AGN jet simulations with high energy flux. Each row represents a different AGN jet simulation. From top to bottom, rows present the precessing kinetic jet, thermal jet, cosmic ray jet, and the fiducial simulation with no-jet for comparison purposes. Each column represents a different parameter constraint applied during the post-processing. In each column, from left to right, we present the logarithm of the overall gas mass density, the O {\footnotesize VIII} surface mass density, the O {\footnotesize VI} surface mass density, the Mg {\footnotesize II} surface mass density, and gas temperature, respectively. The color bars depicted at the bottom of the figure show the intensity of each projection map.  
}
\label{fig:high-energy-xy}
\end{figure*}    

\begin{figure*}[th!]
\centering
\includegraphics[scale=0.34]{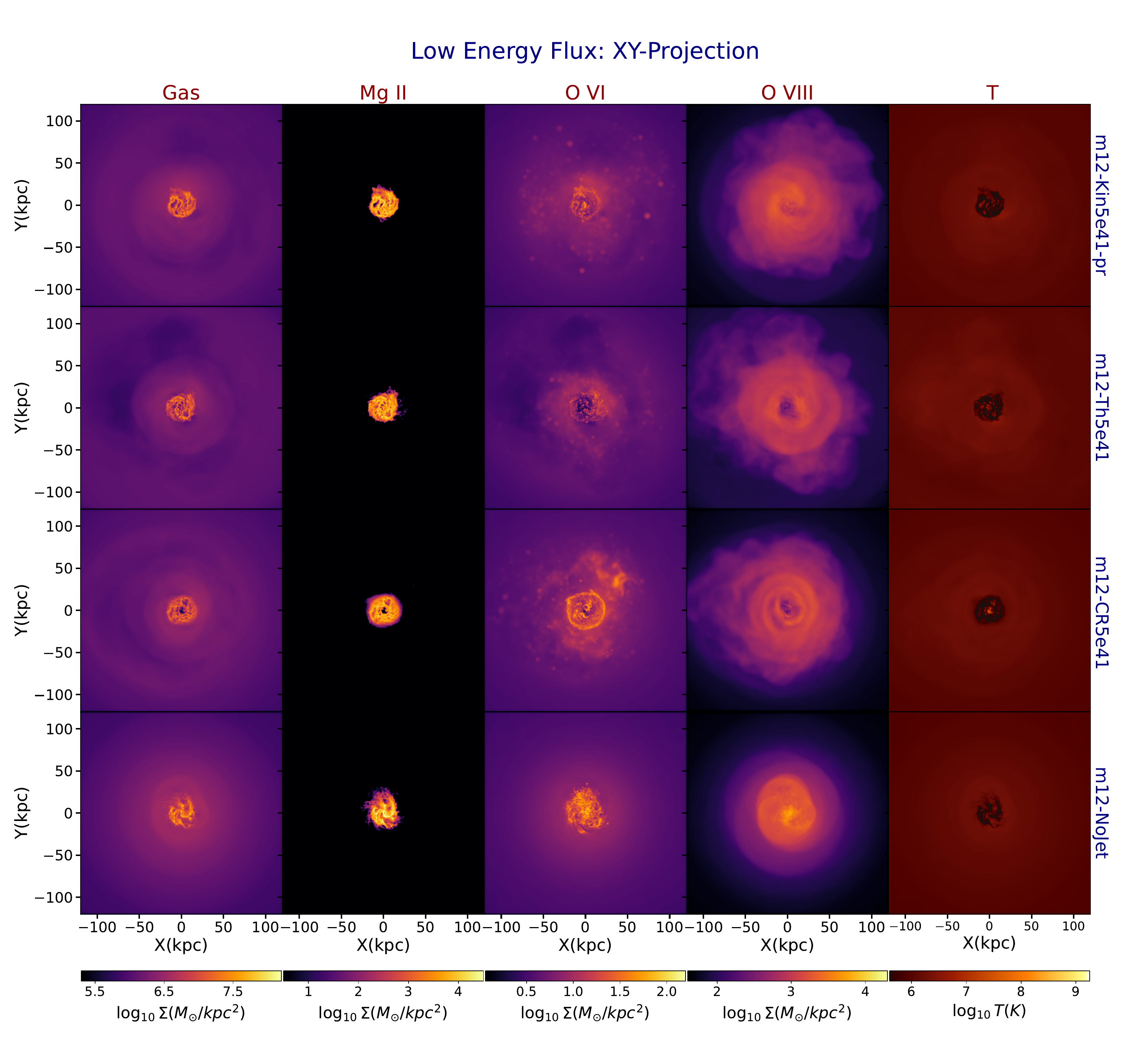} 
\caption{The face-on (XY-projected) view of different AGN jet simulations with low energy flux. Each row represents a different AGN jet simulation. From top to bottom, rows present the precessing kinetic jet, thermal jet, cosmic ray jet, and the fiducial simulation with no-jet for comparison purposes. Each column represents a different parameter constraint applied during the post-processing. In each column, from left to right, we present the logarithm of the overall gas mass density, the O {\footnotesize VIII} surface mass density, the O {\footnotesize VI} surface mass density, the Mg {\footnotesize II} surface mass density, and gas temperature, respectively. The color bars depicted at the bottom of the figure show the intensity of each projection map.  
}
\label{fig:low-energy-xy}
\end{figure*}    

\begin{figure*}[th!]
\centering
\includegraphics[scale=0.34]{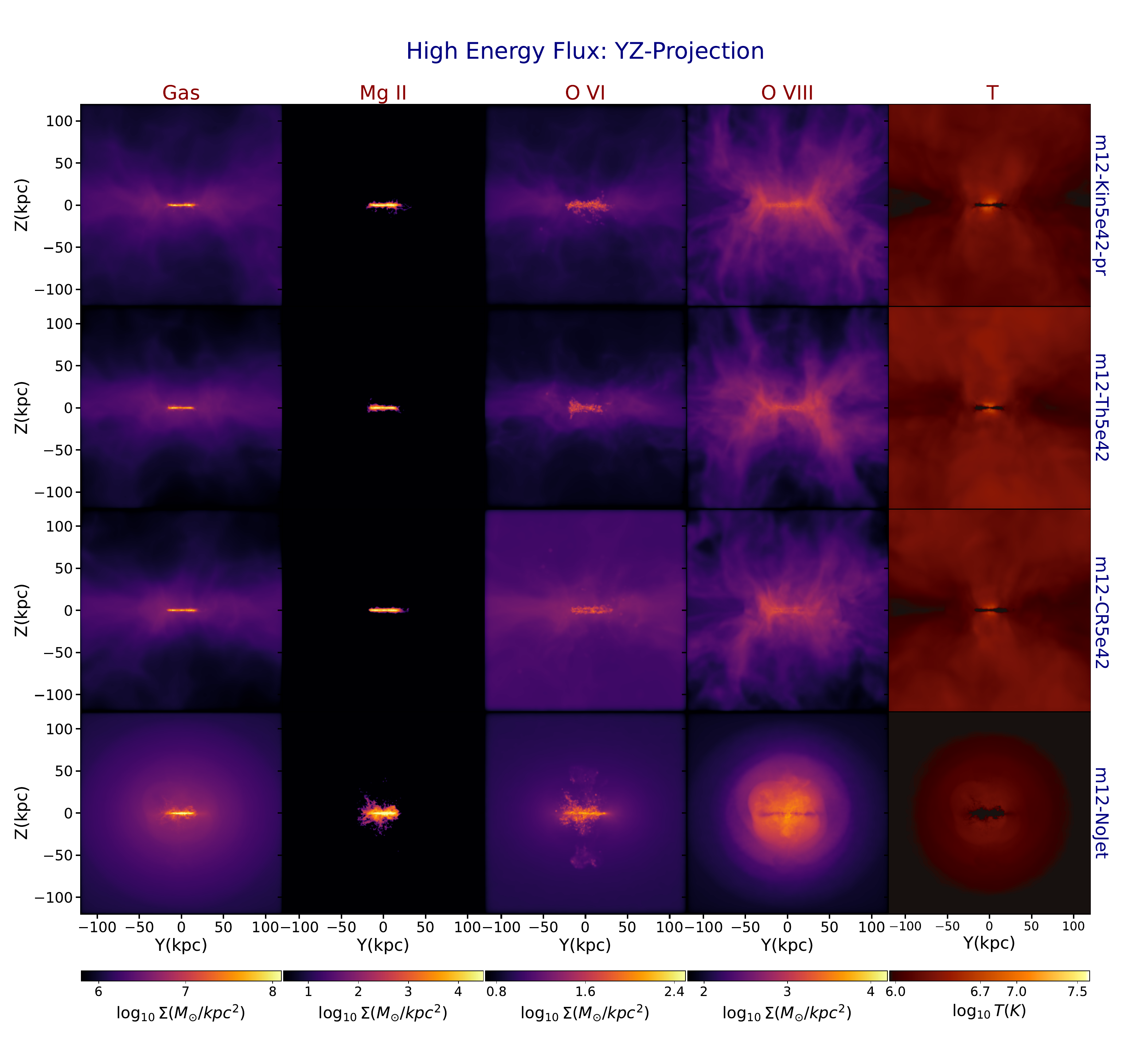}
\caption{The edge-on (YZ-projected) view of different AGN jet simulations with high energy flux. Each row represents a different AGN jet simulation. From top to bottom, rows present the precessing kinetic jet, thermal jet, cosmic ray jet, and the fiducial simulation with no-jet for comparison purposes. Each column represents a different parameter constraint applied during the post-processing. In each column, from left to right, we present the logarithm of the overall gas mass density, the O {\footnotesize VIII} surface mass density, the O {\footnotesize VI} surface mass density, the Mg {\footnotesize II} surface mass density, and gas temperature, respectively. The color bars depicted at the bottom of the figure show the intensity of each projection map. 
}
\label{fig:high-energy-yz}
\end{figure*}    

\begin{figure*}[th!]
\centering
\includegraphics[scale=0.34]{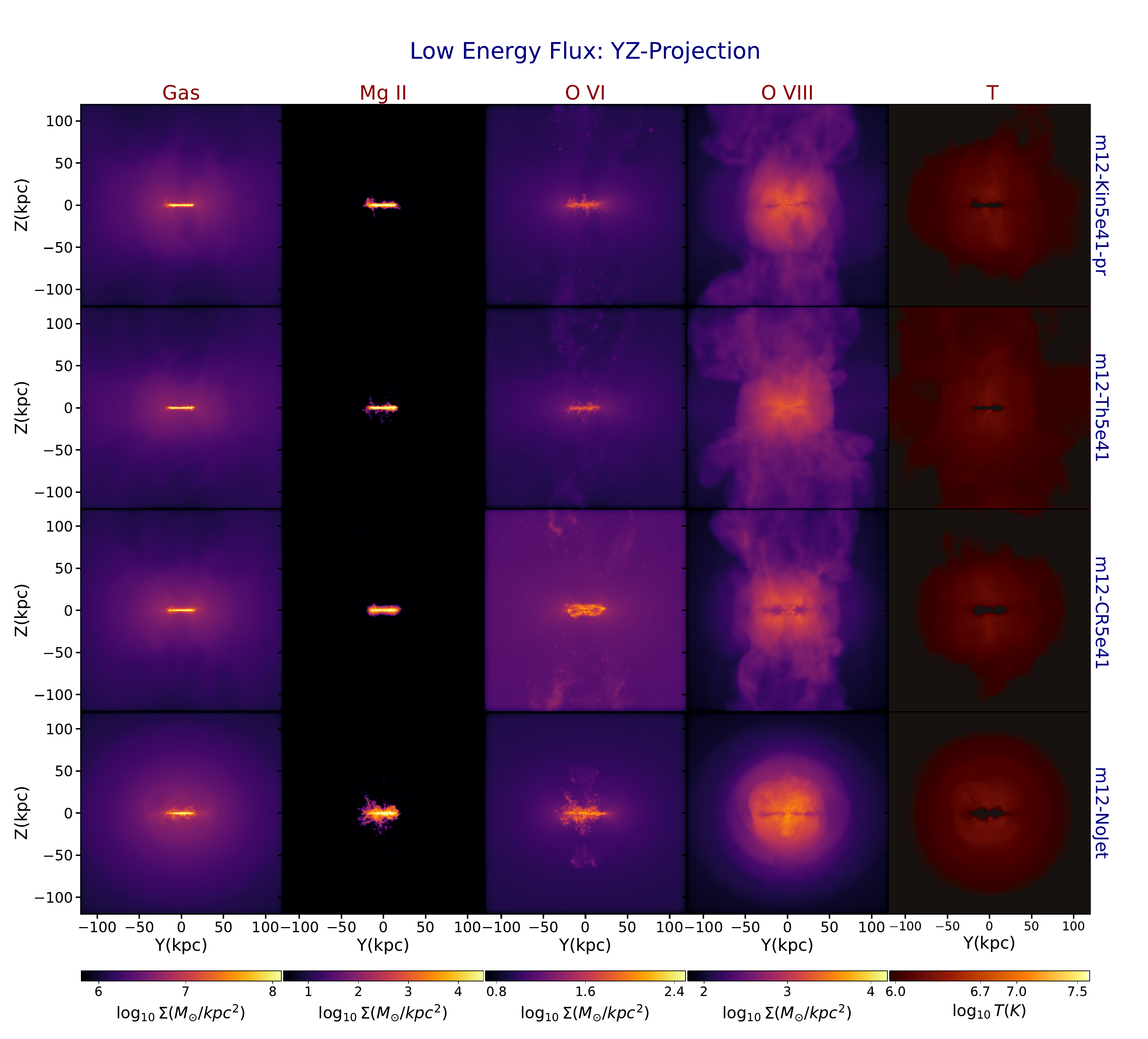}
\caption{The edge-on (YZ-projected) view of different AGN jet simulations with low energy flux. Each row represents a different AGN jet simulation. From top to bottom, rows present the precessing kinetic jet, thermal jet, cosmic ray jet, and the fiducial simulation with no-jet for comparison purposes. Each column represents a different parameter constraint applied during the post-processing. In each column, from left to right, we present the logarithm of the overall gas mass density, the O {\footnotesize VIII} surface mass density, the O {\footnotesize VI} surface mass density, the Mg {\footnotesize II} surface mass density, and gas temperature, respectively. The color bars depicted at the bottom of the figure show the intensity of each projection map.  
}
\label{fig:low-energy-yz}
\end{figure*}    

\subsection{Post-Processing Calculation of Ion Distributions} 
\label{sec:computationalanalysis} 
We employ {\small TRIDENT} \citep{2017ApJ...847...59H} to compute the spatial distribution of three distinct ions— Mg {\small II}, O {\small VI}, O {\small VIII}.  {\small TRIDENT} facilitates the generation of simulated observations within astronomical hydrodynamic simulations, allowing the creation of absorption line spectra and column density maps for ion species not initially present in the simulation outputs. The process involves {\small TRIDENT} initially computing the density of a specific ion within the simulated dataset. This computation uses its ion balance module, which assumes an ionizing source, as described in \cite{2012ApJ...746..125H}. This module assesses if the dataset contains the particular metal element for each cell within the considered domain. If the simulation includes the ion density via the chemistry solver, that data is utilized. Alternatively, {\small TRIDENT} estimates the density of these ions by assuming chemical equilibrium when the specific ion density is not tracked in the simulation.

We undertake a comparative analysis involving a baseline simulation devoid of AGN jet feedback and six distinct AGN jet feedback models, as detailed in Section \ref{sec: AGN-models}. Figure \ref{fig:Jet propagation} illustrates the progression of the jet at various snapshots. Each panel depicts the mass-weighted temperature in the Y-Z projection. Our investigation delves into the distinctive signatures exhibited by various AGN feedback models across a selection of ions at varying ionization states, specifically Mg {\small II}, O {\small VI}, and O {\small VIII}, associated with low, intermediate, and high ionized states, respectively. The strategic selection of ions with significantly divergent ionization energies allows for a comprehensive sampling of each phase of gas present in the ISM and the CGM.

\section{Results}
\label{sec:results}
In this section, we undertake a comprehensive examination of the CGM properties associated with each of the aforementioned jet models. Our numerical approach is divided into two distinct segments. In the initial segment, our attention is directed towards a singular snapshot, offering a detailed insight into the specific characteristics observed. Subsequently, in the second segment, we delve into a thorough investigation of the CGM ionization distribution, conducted at a time-median level, providing a nuanced understanding of the temporal dynamics of the jet. 

\textbf{Single Snapshot Analysis(SSA):} 
We initiate our exploration through the adoption of a Single Snapshot Analysis (SSA) methodology, specifically selecting snapshot $N = 100$, corresponding to $\Delta t =1 \mathrm{Gyr}$ from the starting point of the simulation. As depicted in the right panel of Figure \ref{fig:Jet propagation}, this particular snapshot captures a discernible jet expansion, extending approximately $\sim$ 300 kpc along the $Z$-direction. This configuration enhances the feasibility of scrutinizing and discerning intricate details pertaining to the jet's inherent properties. Movies of individual simulations including all different snapshots are shown at \href{https://www.youtube.com/@nadiaqutob/videos}{here}. Below, we utilize the SSA method to generate Figures \ref{fig:high-energy-xy} through \ref{fig:low-energy-yz}.

\begin{figure*}[th!]
    \centering
    \includegraphics[scale=0.41]{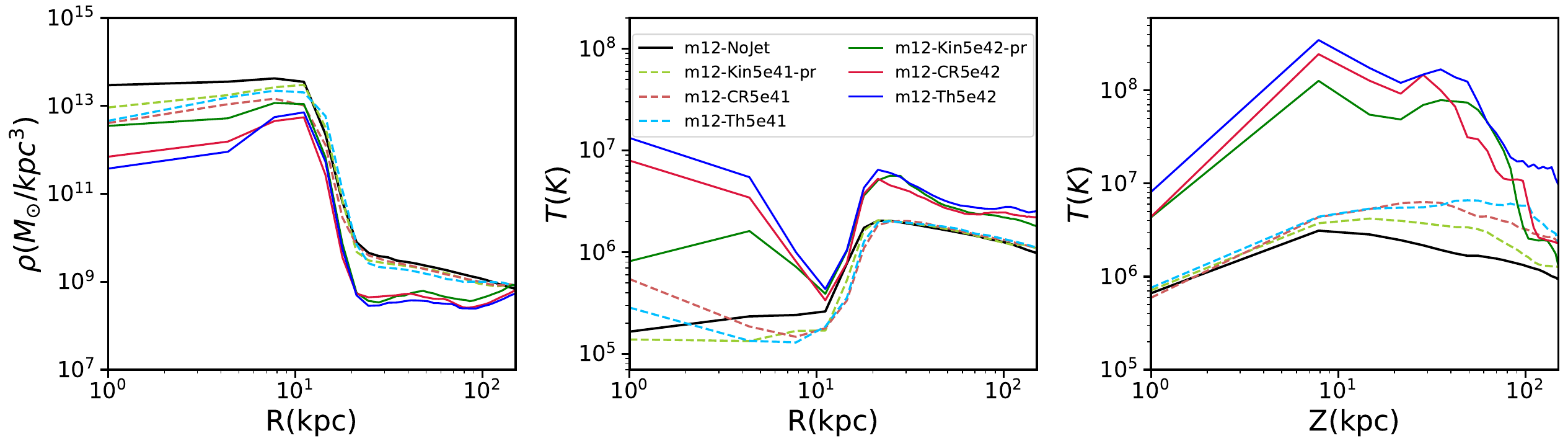}
    \caption{Left panel: Median gas mass-density in the galaxy to distance from the center of the galaxy along the x-axis. Middle Panel: The median temperature vs the radial profile. Right Panel: The median temperature vs the distance from the Z-axis.  Overlaid in each panel we present different simulations using different colored lines. }  
    \label{fig:Temperature-density}
\end{figure*}

\textbf{Multiple Snapshot Analysis (MSA):} Building upon the SSA method, we broaden our investigation to encompass the dynamical evolution of jet properties using a time-median approach, referred to as Multiple Snapshot Analysis (MSA) method. This involves calculating the median values of gas density, mass-weighted temperature, and the mass ratios and densities of our targeted three ions—Mg {\small II}, O {\small VI}, and O {\small VIII}. Below, we utilize the MSA method to generate Figures \ref{fig:Temperature-density} through \ref{fig:mass-density-Jet-Disk}.

\subsection{Jet Cocoon Morphology} \label{sec:jetmorphology} 
Figures \ref{fig:high-energy-xy} and \ref{fig:low-energy-xy} delineate a face-on view (i.e. X-Y projection) of simulations corresponding to high and low energy flux, respectively. In making this figure we have used SSA method. Down rows represent simulations involving precessing-kinetic jet, hot thermal jet, cosmic ray jet, and a simulation with no-jet. In each row, from the left to right, we present the projected density of gas, Mg {\small II}, O {\small VI} and O {\small VIII}, as well as the mass-weighted temperature profiles. Notably, Figure \ref{fig:high-energy-xy} illustrates that the introduction of high energy flux results in a low-density heated region in the ISM surrounding the central AGN. This, in turn, leads to a suppression of the cooling flow, significantly reducing the late-stage gas supply from the ISM. Consequently, an overall decrease in the mass of the ISM is observed, visually manifested as a void in the X-Y projection, attributed to the presence of the jet. It is noteworthy that this central under-dense region is less pronounced in simulations with lower energy flux jets, as depicted in Figure \ref{fig:low-energy-xy}.

Figures \ref{fig:high-energy-yz} and \ref{fig:low-energy-yz} present an edge-on perspective (corresponding to a Y-Z projection) of simulations related to high and low energy flux, respectively, mirroring the structure of Figures \ref{fig:high-energy-xy} and \ref{fig:low-energy-xy}, utilizing SSA method. The edge-on view is instrumental in observing the propagation of the jet, confirming the presence of a pressurized cocoon surrounding it. AGN jet cocoons are characterized by lower density and higher temperature compared to their surroundings. As these jet cocoons propagate and expand laterally, they impart heat to the surrounding CGM and expel gas as an outflow. Consequently, this outflow has a suppressive impact on star formation.

\subsection{Density and Temperature} 
\label{sec:mass}
The first column in Figures \ref{fig:high-energy-xy}-\ref{fig:low-energy-yz} displays the column density distribution of gas within each analyzed galaxy. Mass is concentrated towards the AGN center in the ISM for all simulations, with higher jet fluxes correlating to lower gas densities. A noticeable void appears in the ISM with a sufficiently high energy flux, corresponding to the region where the AGN jet expels energy into the CGM. This effect is due to the AGN jet depositing energy onto the surrounding gas, causing dispersion and the formation of a high-temperature-low-density gas region within and around the jet cocoon, as shown in the 5th column of Figures \ref{fig:high-energy-xy}-\ref{fig:low-energy-yz}. This universal suppressive trend in both ISM and CGM aligns with expectations based on rough pressure equilibrium, where heated regions exhibit lower density and vice versa.

The jet cocoon primarily consists of high-temperature gas, while the disk remains relatively cool, exhibiting a development of a cold gas phase. Notably, at radii of approximately \(12\mathrm{kpc}\), there is an observable ring of cold gas in the equatorial plane, where residual star formation, although significantly suppressed by the jet, continues to take place. Despite this, the sparse nature of the star formation results in the galaxy being considered quenched. Moreover, at farther radii, around \( 200\mathrm{kpc}\), the CGM exhibits a lower mass density compared to the inner disk. These discernible patterns remain consistent across all analyzed simulations.

Figure \ref{fig:Temperature-density} employs the MSA method to calculate the radial profile of the median gas density (left panel) and the median mass-weighted gas temperature (middle panel), along with the Z-dependence of the median mass-weighted gas temperature (right panel). The plot is generated using 60 linear shells covering radii from \(1-200 \mathrm{kpc}\). 

\textbf{In the first two columns}, it is evident that cool gas exhibits higher density in the ISM. Furthermore, simulations with AGN jets, whether low energy flux or no-jet, maintain a cool ISM, while those with higher energy flux induce heating in the inner few kpc, resulting in a decreasing temperature profile towards the ISM edge. Consequently, a direct measurement of the gas temperature profile in the simulation could provide insights into the AGN jet's energy flux. Notably, there is a break in the slope around  \(10 \mathrm{kpc}\) in both density and temperature, corresponding to the ISM to CGM transition, where density drops substantially while temperature increases. Another break occurs at roughly  \(30 \mathrm{kpc}\), marking the cooling radius, within which gas is cooled into multiple phases over a billion years. It is evident that with the same energy flux, thermal and cosmic ray jets are notably more effective in suppressing ISM density. This efficiency could stem from the pressure exerted by cosmic ray and thermal jets, especially when directed near the black hole, proving highly effective in suppressing core density. Interestingly, both cosmic ray and thermal jets demonstrate a more efficient capability in heating the ISM compared to precessing kinetic jets with equivalent energy fluxes. However, as the jet propagates, the kinetic energy within the precessing kinetic jet undergoes eventual shock and thermalization. Consequently, different jet models exhibiting the same energy flux tend to suppress CGM density to a similar extent at larger radii. 

\textbf{In the third column}, it is evident that while low and high flux jets have similar temperature profiles in the inner ISM, they diverge at the ISM edge, with higher flux jet models showing a steep growth and low energy flux AGN models exhibiting a relatively flat profile.

\begin{figure*}[th!]
\centering
\includegraphics[scale=0.35]{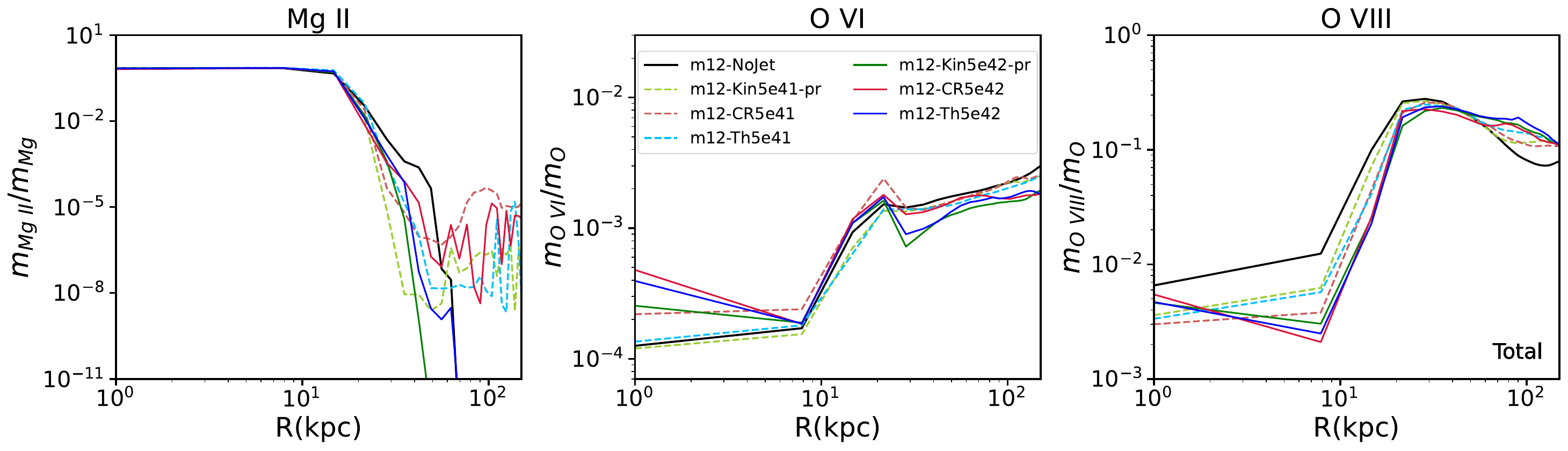}
\includegraphics[scale=0.35]{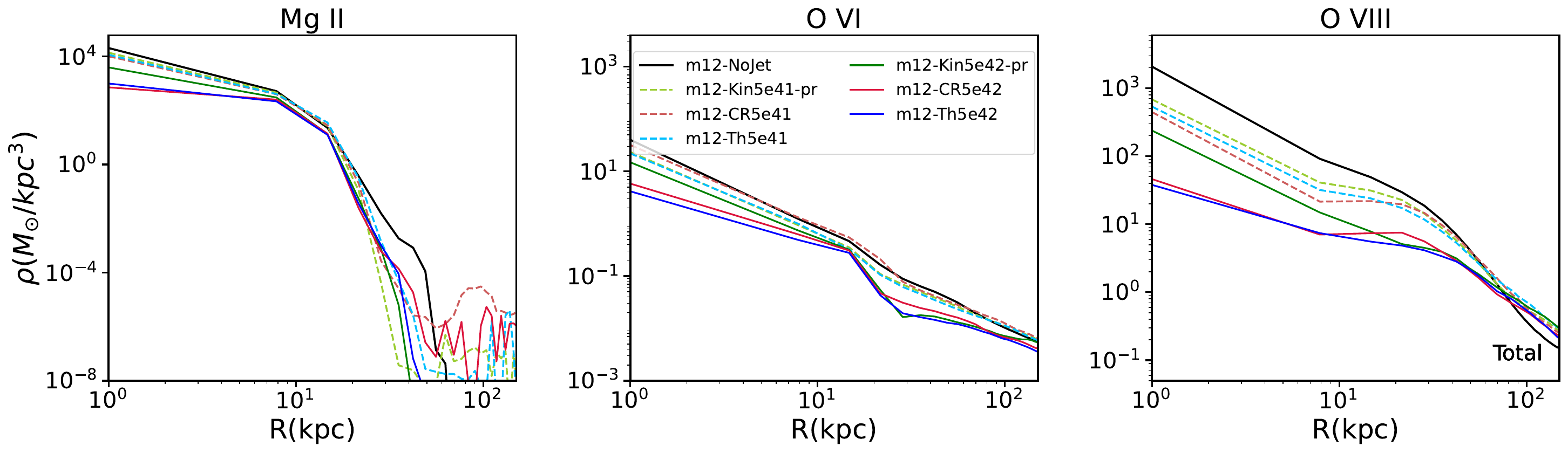}
\caption{The radial distribution of the median of the mass density (top-row) and the ion-to-element mass ratio (bottom-row) from all snapshots for Mg {\footnotesize II} (left panel), O {\footnotesize VI} (middle panel) and O {\footnotesize VIII} (right panel), respectively. }
\label{fig:mass-density-noscatter}
\end{figure*}    

\begin{figure*}[th!]
\centering
\includegraphics[scale=0.35]{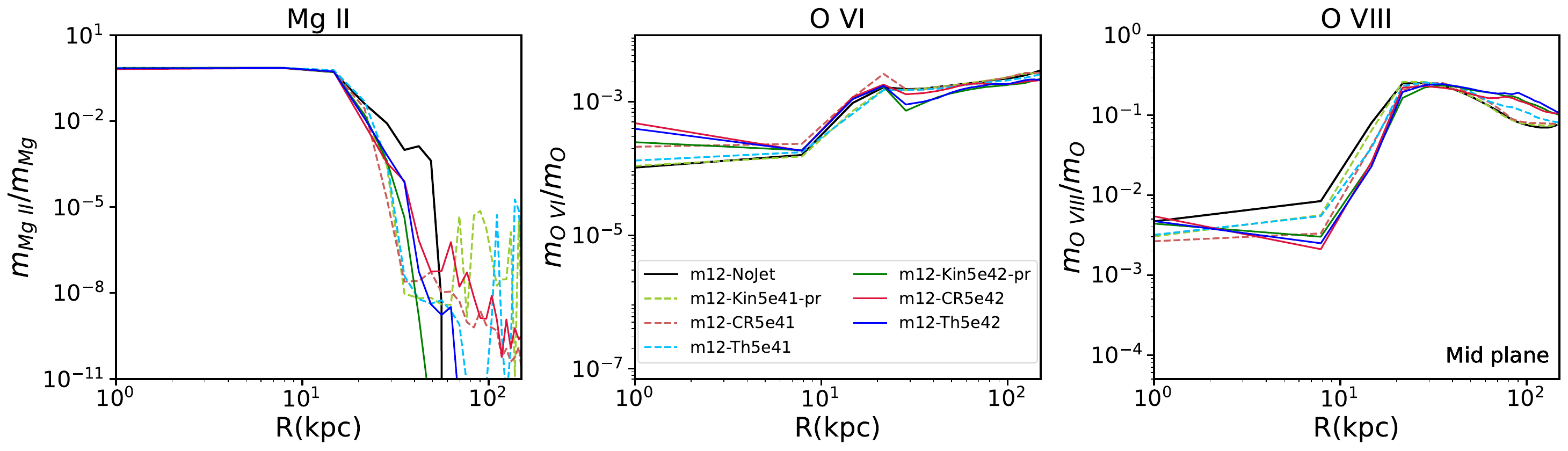}
\includegraphics[scale=0.35]{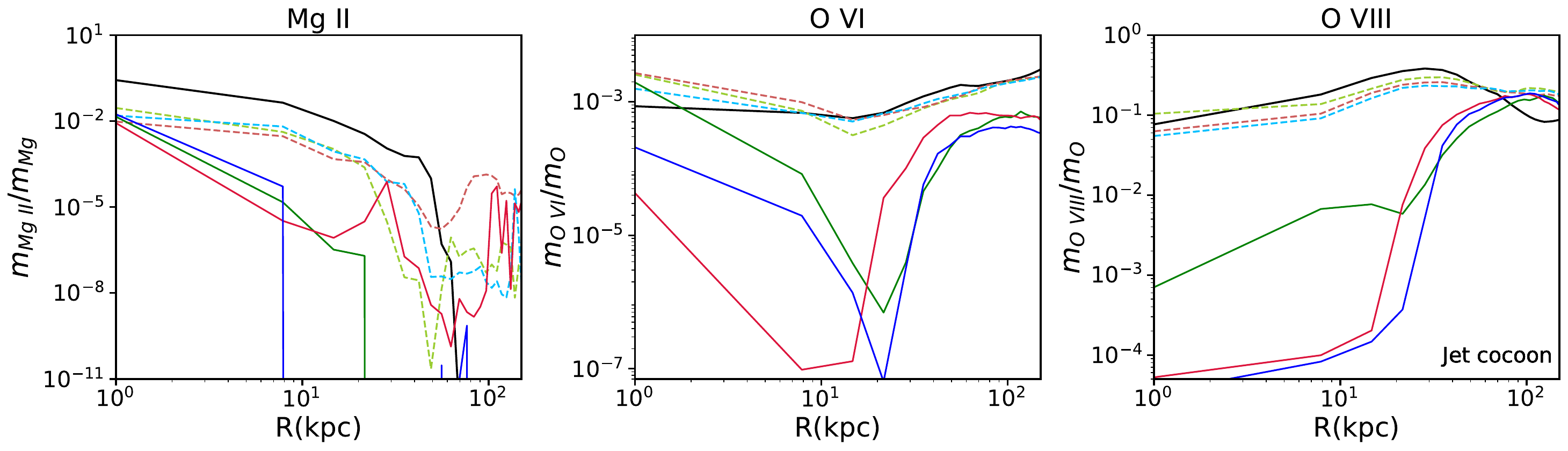}
\caption{The depiction of the mass ratio for Disk (Top panel) and the jet (bottom panel). From the left to right we present the Mg {\footnotesize II}, O {\footnotesize VI} and O {\footnotesize VIII}, respectively. }
\label{fig:mass-ratio-Jet-Disk}
\end{figure*}    

\begin{figure*}[th!]
\centering
\includegraphics[scale=0.35]{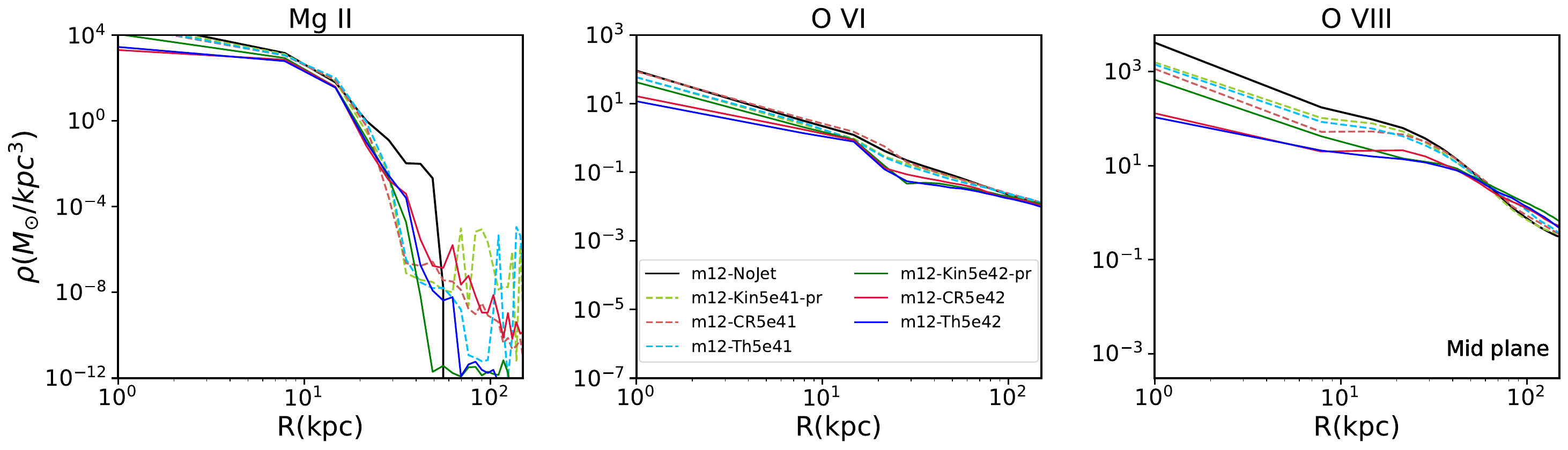}
\includegraphics[scale=0.35]{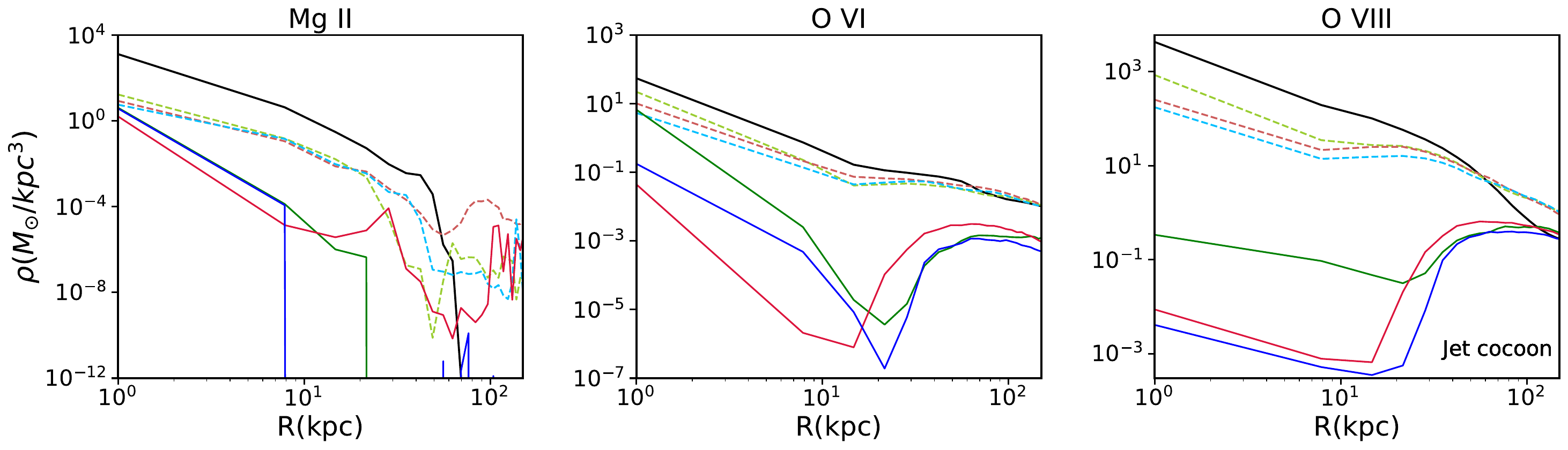}
\caption{The depiction of the mass density for the Disk (Top panel) and the jet (bottom panel). From the left to right we present the Mg {\footnotesize II}, O {\footnotesize VI} and O {\footnotesize VIII}, respectively. }
\label{fig:mass-density-Jet-Disk}
\end{figure*}    

\subsection{General Ion Distributions} \label{sec:ionanalysis} 
In Milky Way-like galaxies, the ISM and CGM typically exhibit temperatures of \(T\gtrsim10^4K\) and $T\geq10^6K$, respectively. To investigate the morphological distribution of ions across the galaxy, we focus on analyzing Mg {\small II}, O {\small VI}, and O {\small VIII}, representing low, intermediate, and high energy ionization states, respectively. These ions cover a wide range of ionization energies, spanning from $10^5 K$ to $10^7$ K, and thus sample various gas phases at different radii. For instance, it's expected that relatively cooler ions like Mg {\small II} are more prevalent in the galactic disk, while hotter ions such as O {\small VI} and O {\small VIII} are more dominant in the CGM. 

Below, we provide a detailed description of the trends in ion mass density and ion-to-element mass ratio distributions for these three different ions (Mg {\small II}, O {\small VI}, and O {\small VIII}). 

\subsection{Mg {\small II}} 
\label{sec:Mg II} 
As previously highlighted, Mg {\small II} exhibits a low ionization temperature, suggesting its prevalence in the cool, dense phase of the ISM within the galaxy. It is revealed from column 4 in Figures \ref{fig:high-energy-xy}-\ref{fig:low-energy-yz} that the majority of the projected surface density for Mg {\small II} in our simulated AGN jets is concentrated within a radius of \(15 \mathrm{kpc}\). This finding aligns with the observations from the first column in Figure \ref{fig:mass-density-noscatter}, where a pronounced decline in Mg {\small II} ion-to-element mass ratio (first row) and mass density (second row) is observed near the ISM-to-CGM transition point.

A comparative analysis of the size of the low-density region, denoted as the hole at the center of the ISM, indicates that higher-energy flux jets (Figure \ref{fig:high-energy-xy}) generate larger holes in the ISM with lower Mg {\small II} density compared to lower-energy flux jets (Figure \ref{fig:low-energy-xy}). In simulations featuring lower-energy flux jets, uniformly higher Mg {\small II} masses within the ISM are observed, as depicted in Figures \ref{fig:low-energy-xy} and \ref{fig:low-energy-yz}. When the jet possesses a sufficiently high energy flux, it not only reduces the overall density at the center but also elevates the gas temperature, leading to further ionization of Mg into higher ions. The dominant role of the first effect is underscored by the similarity in ion-to-element mass ratios for Mg {\small II} in the ISM across different simulations.

Lower energy flux cosmic ray jets exhibit a slightly greater amount of cold gas at larger radii compared to other runs, including the no-jet case. This observation is quantified in Figure \ref{fig:mass-density-noscatter} at radii around \(50-100 \mathrm{kpc}\), and it is consistently illustrated by the presence of sparse clumps with Mg {\small II} at similar radii in Figure \ref{fig:low-energy-yz}. This phenomenon becomes more pronounced for Mg {\small II} due to its lower temperature. The observed behavior aligns with the concept that additional pressure support from cosmic rays can sustain more cold gas at larger radii \citep[e.g.,][]{2020MNRAS.496.4221J,2020ApJ...903...77B,2020MNRAS.492.3465H, 2023MNRAS.521.2477B,2022ApJ...935...69B}.

In Figures \ref{fig:mass-ratio-Jet-Disk} and \ref{fig:mass-density-Jet-Disk}, we present the ion-to-element mass ratio and mass density for Mg {\small II} in the disk (top panel) and jet cocoon (bottom panel) regions, respectively. We define `disk' and `jet cocoon' as the areas located within 45$^\circ$ of the equatorial plane as well as the Z-axis, respectively. The plots distinctly reveal that within the ISM, both the ion-to-element mass ratio (Figure \ref{fig:mass-ratio-Jet-Disk}) and mass density (Figure \ref{fig:mass-density-Jet-Disk}) experience a decline in the jet compared to the disk. As expected, the high energy flux jets exhibit a more pronounced decline in these quantities than the lower flux jets. However, the trend diverges in the CGM, where, with the exception of the cosmic ray jet, high-energy flux jets experience some level of decline of Mg {\small II} in the jet cocoon region, while the low-energy jet models have more sustained  Mg {\small II} in the jet cocoon than in the disk. This is most likely because low-energy jets expel some cooler gas to large radii without significantly heating it up, and the expelled gas can also further induce cooling. Cosmic ray jet can do so even with high energy flux, as we see non-negligible Mg {\small II} extended to over 100 kpc. Mg {\small II} provides a clear depiction of the jet's influence on the cold gas within the ISM, as it exhibits significant suppression within the jet cavity and the hot region surrounding the galactic disk. 

\subsection{O {\small VI}} 
\label{sec:O VI}
O {\small VI} is indicative of a medium-temperature state, making its presence likely in both the ISM and the CGM. This renders O {\small VI} a valuable comparative tool for assessing the impact of various types of jets on multiple galactic regions simultaneously. In line with our observations in mass density and temperature projections, the discernible differences between different AGN jets in the ISM are also reflected in the behavior of O {\small VI}, as depicted in the third column of Figures \ref{fig:high-energy-xy}-\ref{fig:low-energy-yz}. This distinction is further evident when comparing high-energy flux scenarios represented in Figures \ref{fig:high-energy-xy} and \ref{fig:high-energy-yz} with low-energy flux scenarios illustrated in Figures \ref{fig:low-energy-xy} and \ref{fig:low-energy-yz}. Across all jet types, the O {\small VI} mass density is more pronounced and extended in the low-energy flux simulations.

The analysis of the second column in Figure \ref{fig:mass-density-noscatter} reveals that the ion-to-element mass ratio, depicted in the top panel, experiences an increase from the ISM to the CGM, eventually reaching saturation at radii exceeding 50 kpc. This observed trend contrasts with the behavior of Mg {\small II}. The second row in Figure \ref{fig:mass-density-noscatter} illustrates a decline in the mass density of O {\small VI}, albeit with a smoother profile compared to the behavior observed for Mg {\small II}.

\begin{figure*}[th!]
\centering
\includegraphics[scale=0.8]{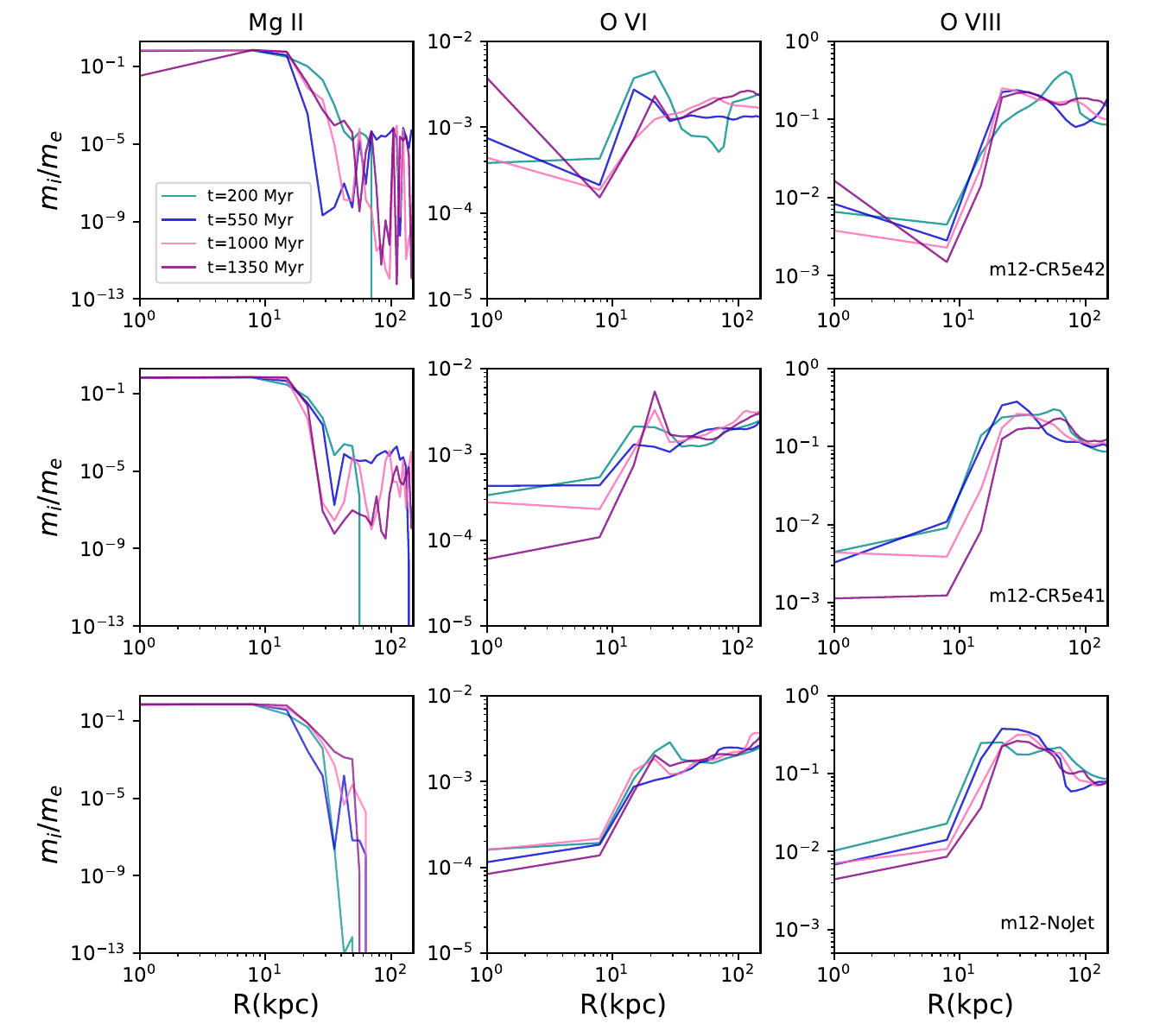}
\caption{The depiction of the mass ratio for cosmic ray jet with high (top panel), low (intermediate panel) energy flux and the default no jet case (bottom panel)for a few different times including t=(200,550,1000,1350) Myr. From left to right, we present the Mg {\footnotesize II}, O {\footnotesize VI}, and O {\footnotesize VIII}, respectively. }
\label{fig:mass-ratio-time}
\end{figure*}    
The second columns in Figures \ref{fig:mass-ratio-Jet-Disk} and \ref{fig:mass-density-Jet-Disk} present the ion-to-element mass ratio and mass density for O {\small VI} in the mid-plain (top panel) and jet cocoon (bottom panel) regions, respectively. It is observed that trends in the mid-plain closely resemble those depicted in Figure \ref{fig:mass-density-noscatter}. In contrast, the jet cocoon region exhibits a distinct behavior, demonstrating a more pronounced decline in both the ion-to-element mass ratio and mass density for high energy flux jet models compared to low energy flux  models. Notably, this decline is predominantly observed for radii in the range between 6-20 kpc. Across all cases, the low-flux jet models exhibit behavior similar to the fiducial model. With an ionizing temperature positioned between the virial temperature and the ISM, O {\small VI} clearly manifests distinctions among identical jets with varying energy fluxes.

\subsection{O {\small VIII}} 
\label{sec:O VIII}
O {\small VIII} possesses the highest ionization energy in comparison to Mg {\small II} and O {\small VI}. Its density distribution, observed in the fourth column of Figures \ref{fig:high-energy-xy}-\ref{fig:low-energy-yz}, is voluminous due to its elevated ionization energy. Remarkably, O {\small VIII} persists even in some of the hottest regions within the analyzed AGN jet cocoons. This characteristic positions O {\small VIII} as a valuable indicator, providing a map of the distribution of hot gas within and around an AGN jet cocoon.

The examination of the third column in Figure \ref{fig:mass-density-noscatter} indicates that the ion-to-element mass ratio, illustrated in the top panel, undergoes an increase from the ISM to the CGM, mirroring the behavior observed for O {\small VI}. The second row in Figure \ref{fig:mass-density-noscatter} further reveals that the range within which O {\small VIII} can sustain itself extends to the virial radius of the galaxy, exhibiting less suppression in the mass density profile compared to Mg {\small II} and O {\small VI}. Notably, different AGN jet models exhibit more pronounced distinctions from one another for O {\small VIII} than for the other ions considered. In simulations with high energy flux, the kinetic jet, within 30 kpc, demonstrates higher density compared to other cases with the same energy flux, attributed to the pressurization of non-kinetic energy at smaller radii, proving more effective in density suppression.

The third columns in Figures \ref{fig:mass-ratio-Jet-Disk} and \ref{fig:mass-density-Jet-Disk} present the ion-to-element mass ratio and mass density for O {\small VIII} in the disk (top panel) and jet cocoon (bottom panel) regions, respectively. It is evident from the third column in Figure \ref{fig:mass-ratio-Jet-Disk} that the ion-to-element mass ratio exhibits an increasing profile for both the disk and jet cocoon, contrasting with the behavior observed for O {\small VI}. However, an analysis of the third column in Figure \ref{fig:mass-density-Jet-Disk} indicates that the density profile for O {\small VI} and O {\small VIII} is consistent in both the disk and jet cocoon.

\section{Discussion}
\label{sec: discussion}

\subsection{Time-Evolution}\label{sec: Time-evolution} 
Figure \ref{fig:mass-ratio-time} presents the evolution of the ion to element mass ratio for cosmic ray jet with high (top panel) and low (bottom panel) energy flux at a few different snapshots, as an example of the range of value through the simulation time.

A few trends could be observed. Initially, Mg {\small II} is present only in the galactic center, where the cold ISM is located. The jet mixes part of the ISM into larger radii, hence increasing Mg {\small II} at larger radii over the first 500 Myr. Meanwhile, the high-energy flux CR jet also heats up the gas, resulting in a later drop of Mg {\small II} at large radii. On the other hand, a lower energy flux cosmic ray jet can sustain a higher overall Mg {\small II} at large radii over an extended period of time.

The O {\small VI} and O {\small VIII} evolve differently over time compared to Mg {\small II}. As the ISM gets heated up by the high-energy flux jet, the O {\small VI} and O {\small VIII} become more prominent at small radii. A lower energy flux cosmic ray jet does not heat up the ISM as much.

\subsection{Jet vs No Jet Comparison}\label{sec: jet vs no jet comparison} 
The no-jet simulation served as the fiducial reference model for our AGN jet simulations. In comparison to each of the other six types of jet simulations, the no-jet simulation exhibited significantly less structure at large radii. Additionally, the no-jet runs displayed a higher concentration of mass density toward the center of the galaxy, attributed to the absence of a jet pushing gas outward.

\subsection{High Energy vs Low Energy Jet Comparison}\label{sec: jet comparison} 
Low-energy jets exhibit lower efficiency in launching extended cocoons compared to their high-energy counterparts. Consequently, there is a significantly higher O {\small VI} and O {\small VIII} column density in the jet direction. Jets with higher fluxes demonstrate a notable ability to suppress gas density, heat the gas, and reduce the density of all considered ions.

\subsection{Different Jets with a Similar Energy Flux}\label{sec: jet comparison} 
In comparisons among jets with the same energy flux, non-kinetic energy jets demonstrate pressurization at smaller radii, resulting in a slightly more effective suppression of core densities, as depicted in Figure \ref{fig:Temperature-density}, across all considered ions. However, as we extend our analysis to larger radii, these differences diminish, as most of the energy is pressurized irrespective of its form.

Cosmic ray jets play a pivotal role in providing additional pressure support in the CGM, allowing for the maintenance of more extended regions of cool and warm gas, notably observed in Mg {\small II} and O {\small VI} as depicted in Figures \ref{fig:mass-density-noscatter}-\ref{fig:mass-density-Jet-Disk}. This significance is particularly pronounced in lower energy flux cases. Nevertheless, as the energy flux increases, the impact of strong gas expulsion diminishes the prominence of this cosmic ray pressure support.

\subsection{Limitation of the Model} \label{conclusion}
We conducted a study utilizing isolated galaxy simulations within the Milky Way mass range, incorporating various types of constant flux jets. The simulations excluded black hole accretion to simulate passive AGN feedback. Therefore, our findings should be interpreted as an average over a limited duration of 1.5 Gyr, without accounting for episodic duty cycles. Throughout the simulation period, all runs consistently maintained CGM density levels, ranging from the assumption of no missing baryons to the actual observed Milky Way CGM density (Su et al. in prep).

The fixed flux AGN jets were intentionally tailored to suppress star formation in Milky Way-mass galaxies within our simulations, leading to minimal star formation in the majority of the runs. Consequently, comparisons with actively star-forming galaxies within the same halo mass range may not yield equivalent results, as the design of the fixed flux AGN jets prioritized the quenching of star formation.

Our isolated galaxy simulations deliberately excluded any cosmological context, such as satellite interactions or mergers, which could serve as significant sources of cool gas in the CGM and interact non-linearly with AGN jets. The exploration of these complexities has been reserved for future studies.

The influence of cosmic ray pressure support becomes apparent in the augmented presence of warm and cold gas at larger radii. Nevertheless, it's important to note that the cold clumps observed in the morphological plots for Mg {\small II} and O {\small VI} lack sufficient resolution, resulting in several point-like structures. The resolution of these cold structures within heated regions has been a persistent challenge in simulations, potentially necessitating the incorporation of multi-phase subgrid feedback models \citep{2023arXiv230107116S} and refined ISM models \citep{2023MNRAS.519.3011W} to effectively address this issue.

We wish to highlight that our use of {\small TRIDENT} for calculating various ion column densities did not include local ionizing sources such as stars or AGN. Our analysis was solely based on the ionizing influence of the z=0 ultraviolet background, as detailed in \cite{2012ApJ...746..125H}. This limitation is particularly relevant for the ionization state of lower ions like Mg {\small II} and O {\small VI} within the ISM at distances less than 15 kpc \citep{2013ApJ...765...89S}, where ionization is more likely to be dominated by stellar sources. In the context of AGN, our focus was on the radio mode of feedback rather than the quasar mode, which suggests a low ionization contribution from AGN. Nonetheless, it is crucial to acknowledge that our approach does encompass the variations in the ionization state resulting from changes in the gas's thermal dynamical properties. We left the inclusion of the local ionizing source for future study.

\subsection{Future Work}  \label{sec: observations} 
In our analysis, we observe that the column densities presented in Figures \ref{fig:high-energy-xy} - \ref{fig:low-energy-yz} are generally consistent with existing observational data, specifically in studies of O {\small VI} and Mg {\small II} (see, for instance, \citealt{2016ApJ...833...54W, 2023MNRAS.524..512Q}). This consistency is also evident when compared with results from other simulation studies \citep[e.g.,][]{2017MNRAS.466.3810F}. However, a comprehensive comparison that encompasses both observational and theoretical models \citep[for example,][] {2023arXiv230600092S} is beyond the scope of this paper and is designated for future research endeavors.

While the primary focus of this initial study centered on three prominent ions covering a range from low to high ionization states, our forthcoming work (Emami et al., in prep) will extend this analysis to encompass other ions, including the behavior of all ions for each element. This expanded scope will enable a clearer understanding of the conversion from one ionic state to another. Our goal is to compile a list of ions exhibiting more promising observational signatures in terms of their radial distribution, equivalent width, and spectra. These results can then be directly compared to current and future observations spanning wavelengths from the UV to the X-ray. Notably, a comparison of resonant absorption lines due to Ca II and Na I will be particularly crucial, as these lines probe the neutral phase of the outflow and can be scrutinized in detail, even at high redshift, with instruments like JWST \citep{Belli_2023, Deugenio_2023}.

\section{Conclusion}\label{conc}
In this manuscript, we conducted a comprehensive study on the distinctive characteristics of various AGN jet models designed to quench Milky Way-like halos in isolated galaxy simulations, as previously detailed in works by \citet{2018MNRAS.473L.111S,2019MNRAS.487.4393S,2021MNRAS.507..175S}. The investigation focused on the morphology of key ions—Mg {\small II}, O {\small VI}, and O {\small VIII}—representing low, intermediate, and high ionization states, respectively. This selection facilitated a comparative analysis of these ions across different jet models to elucidate the energy distribution within AGN jets. The considered jet models encompassed a precessing kinetic jet, a hot thermal jet, and a cosmic ray jet, each featuring two energy fluxes within the range of $5\times 10^{41}-5\times 10^{42}$ erg s$^{-1}$. While all three jet models effectively suppressed star formation and halted the cooling flow at sufficiently high energy fluxes, they displayed noteworthy differences in ion distributions. The key findings are summarized as follows:

A critical aspect of our analysis involves comparing different jet models to a no-jet simulation, allowing for the identification of discernible trends. The radial profile of mass density and the distribution of all ions exhibit more pronounced suppression in high-energy flux jets, particularly in cases involving thermal and cosmic ray jets. This suppression materializes as a hole, indicative of a low-density region within the ISM. Notably, the appearance of this hole is delayed in models featuring a precessing kinetic jet.

Mg {\small II} predominantly confines itself to the galactic disk and is rarely observed in the vicinity of AGN jet cocoons, except in instances of lower-energy-flux cosmic ray jets. In this scenario, cosmic ray pressure support results in the formation of cool clumps containing Mg {\small II} and O {\small VI}, extending up to a radius of 
100kpc. O {\small VI} is more conspicuous in low-energy flux jets, particularly in the case of cosmic ray jets, where cosmic ray pressure support leads to an extended distribution of O {\small VI} in the inner CGM. O {\small VIII} emerges prominently in all tested jet cocoon types, especially at the shock front resulting from gas compression induced by the jets. The primary convergence point among jets for O {\small VIII} occurs at very large radii, approximately 100kpc. Notably, breaks in the profiles of ion-to-element mass fractions for O {\small VI} and O {\small VIII} are observed at the transition point from the ISM to the CGM at a radius of 10kpc.

\section*{Data Availability}
Data directly corresponding to this manuscript and the figures are available to be shared on reasonable request from the corresponding author.

\section*{Acknowledgements} \label{sec:acknowledgements}
It is a great pleasure to thank Daniel Eisenstein and Matthew Ashby for very fruitful conversation that greatly improved the quality of this manuscript. Nadia Qutob acknowledges the Harvard Smithsonian SAO REU program which provided funding for this summer research. The SAO REU program is funded in part by the National Science Foundation REU and Department of Defense ASSURE programs under NSF Grant no. AST-2050813, and by the Smithsonian Institution. Razieh Emami acknowledges the support from grant numbers 21-atp21-0077, NSF AST-1816420, and HST-GO-16173.001-A as well as the Institute for Theory and Computation at the Center for Astrophysics. Kung-Yi Su acknowledges support from the Black Hole Initiative at Harvard University, which is funded by grants from the John Templeton Foundation and the Gordon and Betty Moore Foundation, and acknowledges ACCESS allocations TG-PHY220027 and  TG-PHY220047 and Frontera allocation AST22010. Kung-Yi Su also extends thanks for all the discussions with the FIRE, SMAUG, and LtU collaborations.

\bibliography{references}

\begin{thebibliography}{}
\expandafter\ifx\csname natexlab\endcsname\relax\def\natexlab#1{#1}\fi
\providecommand{\url}[1]{\href{#1}{#1}}
\providecommand{\dodoi}[1]{doi:~\href{http://doi.org/#1}{\nolinkurl{#1}}}
\providecommand{\doeprint}[1]{\href{http://ascl.net/#1}{\nolinkurl{http://ascl.net/#1}}}
\providecommand{\doarXiv}[1]{\href{https://arxiv.org/abs/#1}{\nolinkurl{https://arxiv.org/abs/#1}}}

\bibitem[{{Belli} {et~al.}(2023){Belli}, {Park}, {Davies}, {Mendel}, {Johnson},
  {Conroy}, {Benton}, {Bugiani}, {Emami}, {Leja}, {Li}, {Maheson}, {Mathews},
  {Naidu}, {Nelson}, {Tacchella}, {Terrazas}, \& {Weinberger}}]{Belli_2023}
{Belli}, S., {Park}, M., {Davies}, R.~L., {et~al.} 2023, arXiv e-prints,
  arXiv:2308.05795, \dodoi{10.48550/arXiv.2308.05795}

\bibitem[{{Bourne} \& {Sijacki}(2017)}]{2017MNRAS.472.4707B}
{Bourne}, M.~A., \& {Sijacki}, D. 2017, \mnras, 472, 4707,
  \dodoi{10.1093/mnras/stx2269}

\bibitem[{{Butsky} {et~al.}(2020){Butsky}, {Fielding}, {Hayward}, {Hummels},
  {Quinn}, \& {Werk}}]{2020ApJ...903...77B}
{Butsky}, I.~S., {Fielding}, D.~B., {Hayward}, C.~C., {et~al.} 2020, \apj, 903,
  77, \dodoi{10.3847/1538-4357/abbad2}

\bibitem[{{Butsky} {et~al.}(2023){Butsky}, {Nakum}, {Ponnada}, {Hummels}, {Ji},
  \& {Hopkins}}]{2023MNRAS.521.2477B}
{Butsky}, I.~S., {Nakum}, S., {Ponnada}, S.~B., {et~al.} 2023, \mnras, 521,
  2477, \dodoi{10.1093/mnras/stad671}

\bibitem[{{Butsky} {et~al.}(2022){Butsky}, {Werk}, {Tchernyshyov}, {Fielding},
  {Breneman}, {Piacitelli}, {Quinn}, {Sanchez}, {Cruz}, {Hummels}, {Burchett},
  \& {Tremmel}}]{2022ApJ...935...69B}
{Butsky}, I.~S., {Werk}, J.~K., {Tchernyshyov}, K., {et~al.} 2022, \apj, 935,
  69, \dodoi{10.3847/1538-4357/ac7ebd}

\bibitem[{Chan {et~al.}(2019)Chan, Kere{\v{s} }, Hopkins, Quataert, Su,
  Hayward, \& Faucher-Gigu{\`{e}}re}]{Chan_2019}
Chan, T.~K., Kere{\v{s} }, D., Hopkins, P.~F., {et~al.} 2019, Monthly Notices
  of the Royal Astronomical Society, 488, 3716, \dodoi{10.1093/mnras/stz1895}

\bibitem[{{D'Eugenio} {et~al.}(2023){D'Eugenio}, {Perez-Gonzalez}, {Maiolino},
  {Scholtz}, {Perna}, {Circosta}, {Uebler}, {Arribas}, {Boeker}, {Bunker},
  {Carniani}, {Charlot}, {Chevallard}, {Cresci}, {Curtis-Lake}, {Jones},
  {Kumari}, {Lamperti}, {Looser}, {Parlanti}, {Rix}, {Robertson}, {Rodriguez
  Del Pino}, {Tacchella}, {Venturi}, \& {Willott}}]{Deugenio_2023}
{D'Eugenio}, F., {Perez-Gonzalez}, P., {Maiolino}, R., {et~al.} 2023, arXiv
  e-prints, arXiv:2308.06317, \dodoi{10.48550/arXiv.2308.06317}

\bibitem[{{Eisenreich} {et~al.}(2017){Eisenreich}, {Naab}, {Choi}, {Ostriker},
  \& {Emsellem}}]{2017MNRAS.468..751E}
{Eisenreich}, M., {Naab}, T., {Choi}, E., {Ostriker}, J.~P., \& {Emsellem}, E.
  2017, \mnras, 468, 751, \dodoi{10.1093/mnras/stx473}

\bibitem[{{Fabian} {et~al.}(1994){Fabian}, {Arnaud}, {Bautz}, \&
  {Tawara}}]{1994ApJ...436L..63F}
{Fabian}, A.~C., {Arnaud}, K.~A., {Bautz}, M.~W., \& {Tawara}, Y. 1994, \apjl,
  436, L63, \dodoi{10.1086/187633}

\bibitem[{{Fielding} {et~al.}(2017){Fielding}, {Quataert}, {McCourt}, \&
  {Thompson}}]{2017MNRAS.466.3810F}
{Fielding}, D., {Quataert}, E., {McCourt}, M., \& {Thompson}, T.~A. 2017,
  \mnras, 466, 3810, \dodoi{10.1093/mnras/stw3326}

\bibitem[{{Gaspari} {et~al.}(2012){Gaspari}, {Ruszkowski}, \&
  {Sharma}}]{2012ApJ...746...94G}
{Gaspari}, M., {Ruszkowski}, M., \& {Sharma}, P. 2012, \apj, 746, 94,
  \dodoi{10.1088/0004-637X/746/1/94}

\bibitem[{{Gaspari} \& {S{\c a}dowski}(2017)}]{2017ApJ...837..149G}
{Gaspari}, M., \& {S{\c a}dowski}, A. 2017, \apj, 837, 149,
  \dodoi{10.3847/1538-4357/aa61a3}

\bibitem[{{Haardt} \& {Madau}(2012)}]{2012ApJ...746..125H}
{Haardt}, F., \& {Madau}, P. 2012, \apj, 746, 125,
  \dodoi{10.1088/0004-637X/746/2/125}

\bibitem[{{H{\"a}ring} \& {Rix}(2004)}]{2004ApJ...604L..89H}
{H{\"a}ring}, N., \& {Rix}, H.-W. 2004, \apjl, 604, L89, \dodoi{10.1086/383567}

\bibitem[{{Hernquist}(1990)}]{1990ApJ...356..359H}
{Hernquist}, L. 1990, \apj, 356, 359, \dodoi{10.1086/168845}

\bibitem[{{Hopkins}(2015)}]{2015MNRAS.450...53H}
{Hopkins}, P.~F. 2015, \mnras, 450, 53, \dodoi{10.1093/mnras/stv195}

\bibitem[{{Hopkins}(2016)}]{2016MNRAS.462..576H}
---. 2016, \mnras, 462, 576, \dodoi{10.1093/mnras/stw1578}

\bibitem[{{Hopkins}(2017)}]{2017MNRAS.466.3387H}
---. 2017, \mnras, 466, 3387, \dodoi{10.1093/mnras/stw3306}

\bibitem[{{Hopkins} {et~al.}(2013){Hopkins}, {Narayanan}, \&
  {Murray}}]{2013MNRAS.432.2647H}
{Hopkins}, P.~F., {Narayanan}, D., \& {Murray}, N. 2013, \mnras, 432, 2647,
  \dodoi{10.1093/mnras/stt723}

\bibitem[{{Hopkins} \& {Raives}(2016)}]{2016MNRAS.455...51H}
{Hopkins}, P.~F., \& {Raives}, M.~J. 2016, \mnras, 455, 51,
  \dodoi{10.1093/mnras/stv2180}

\bibitem[{{Hopkins} {et~al.}(2018{\natexlab{a}}){Hopkins}, {Wetzel},
  {Kere{\v{s}}}, {Faucher-Gigu{\`e}re}, {Quataert}, {Boylan-Kolchin}, {Murray},
  {Hayward}, \& {El-Badry}}]{2018MNRAS.477.1578H}
{Hopkins}, P.~F., {Wetzel}, A., {Kere{\v{s}}}, D., {et~al.} 2018{\natexlab{a}},
  \mnras, 477, 1578, \dodoi{10.1093/mnras/sty674}

\bibitem[{{Hopkins} {et~al.}(2018{\natexlab{b}}){Hopkins}, {Wetzel},
  {Kere{\v{s}}}, {Faucher-Gigu{\`e}re}, {Quataert}, {Boylan-Kolchin}, {Murray},
  {Hayward}, {Garrison-Kimmel}, {Hummels}, {Feldmann}, {Torrey}, {Ma},
  {Angl{\'e}s-Alc{\'a}zar}, {Su}, {Orr}, {Schmitz}, {Escala}, {Sanderson},
  {Grudi{\'c}}, {Hafen}, {Kim}, {Fitts}, {Bullock}, {Wheeler}, {Chan},
  {Elbert}, \& {Narayanan}}]{2018MNRAS.480..800H}
---. 2018{\natexlab{b}}, \mnras, 480, 800, \dodoi{10.1093/mnras/sty1690}

\bibitem[{{Hopkins} {et~al.}(2020){Hopkins}, {Chan}, {Garrison-Kimmel}, {Ji},
  {Su}, {Hummels}, {Kere{\v{s}}}, {Quataert}, \&
  {Faucher-Gigu{\`e}re}}]{2020MNRAS.492.3465H}
{Hopkins}, P.~F., {Chan}, T.~K., {Garrison-Kimmel}, S., {et~al.} 2020, \mnras,
  492, 3465, \dodoi{10.1093/mnras/stz3321}

\bibitem[{{Hummels} {et~al.}(2017){Hummels}, {Smith}, \&
  {Silvia}}]{2017ApJ...847...59H}
{Hummels}, C.~B., {Smith}, B.~D., \& {Silvia}, D.~W. 2017, \apj, 847, 59,
  \dodoi{10.3847/1538-4357/aa7e2d}

\bibitem[{{Ji} {et~al.}(2020){Ji}, {Chan}, {Hummels}, {Hopkins}, {Stern},
  {Kere{\v{s}}}, {Quataert}, {Faucher-Gigu{\`e}re}, \&
  {Murray}}]{2020MNRAS.496.4221J}
{Ji}, S., {Chan}, T.~K., {Hummels}, C.~B., {et~al.} 2020, \mnras, 496, 4221,
  \dodoi{10.1093/mnras/staa1849}

\bibitem[{{Kroupa}(2002)}]{2002Sci...295...82K}
{Kroupa}, P. 2002, Science, 295, 82, \dodoi{10.1126/science.1067524}

\bibitem[{{Leitherer} {et~al.}(1999){Leitherer}, {Schaerer}, {Goldader},
  {Delgado}, {Robert}, {Kune}, {de Mello}, {Devost}, \&
  {Heckman}}]{1999ApJS..123....3L}
{Leitherer}, C., {Schaerer}, D., {Goldader}, J.~D., {et~al.} 1999, \apjs, 123,
  3, \dodoi{10.1086/313233}

\bibitem[{{Li} \& {Bryan}(2014)}]{2014ApJ...789...54L}
{Li}, Y., \& {Bryan}, G.~L. 2014, \apj, 789, 54,
  \dodoi{10.1088/0004-637X/789/1/54}

\bibitem[{{Li} {et~al.}(2018){Li}, {Yuan}, {Mo}, {Yoon}, {Gan}, {Ho}, {Wang},
  {Ostriker}, \& {Ciotti}}]{2018ApJ...866...70L}
{Li}, Y.-P., {Yuan}, F., {Mo}, H., {et~al.} 2018, \apj, 866, 70,
  \dodoi{10.3847/1538-4357/aade8b}

\bibitem[{{Martizzi} {et~al.}(2019){Martizzi}, {Quataert},
  {Faucher-Gigu{\`e}re}, \& {Fielding}}]{2019MNRAS.483.2465M}
{Martizzi}, D., {Quataert}, E., {Faucher-Gigu{\`e}re}, C.-A., \& {Fielding}, D.
  2019, \mnras, 483, 2465, \dodoi{10.1093/mnras/sty3273}

\bibitem[{{Navarro} {et~al.}(1996){Navarro}, {Frenk}, \&
  {White}}]{1996ApJ...462..563N}
{Navarro}, J.~F., {Frenk}, C.~S., \& {White}, S. D.~M. 1996, \apj, 462, 563,
  \dodoi{10.1086/177173}

\bibitem[{{Pellegrini} {et~al.}(2018){Pellegrini}, {Ciotti}, {Negri}, \&
  {Ostriker}}]{2018ApJ...856..115P}
{Pellegrini}, S., {Ciotti}, L., {Negri}, A., \& {Ostriker}, J.~P. 2018, \apj,
  856, 115, \dodoi{10.3847/1538-4357/aaae07}

\bibitem[{{Qu} {et~al.}(2023){Qu}, {Chen}, {Rudie}, {Johnson}, {Zahedy},
  {DePalma}, {Boettcher}, {Cantalupo}, {Chen}, {Cooksey},
  {Faucher-Gigu{\`e}re}, {Li}, {Lopez}, {Schaye}, \&
  {Simcoe}}]{2023MNRAS.524..512Q}
{Qu}, Z., {Chen}, H.-W., {Rudie}, G.~C., {et~al.} 2023, \mnras, 524, 512,
  \dodoi{10.1093/mnras/stad1886}

\bibitem[{{Ruszkowski} {et~al.}(2017){Ruszkowski}, {Yang}, \&
  {Reynolds}}]{2017ApJ...844...13R}
{Ruszkowski}, M., {Yang}, H.-Y.~K., \& {Reynolds}, C.~S. 2017, \apj, 844, 13,
  \dodoi{10.3847/1538-4357/aa79f8}

\bibitem[{{Shen} {et~al.}(2013){Shen}, {Madau}, {Guedes}, {Mayer}, {Prochaska},
  \& {Wadsley}}]{2013ApJ...765...89S}
{Shen}, S., {Madau}, P., {Guedes}, J., {et~al.} 2013, \apj, 765, 89,
  \dodoi{10.1088/0004-637X/765/2/89}

\bibitem[{{Smith} {et~al.}(2023){Smith}, {Fielding}, {Bryan}, {Kim},
  {Ostriker}, {Somerville}, {Stern}, {Su}, {Weinberger}, {Hu}, {Forbes},
  {Hernquist}, {Burkhart}, \& {Li}}]{2023arXiv230107116S}
{Smith}, M.~C., {Fielding}, D.~B., {Bryan}, G.~L., {et~al.} 2023, arXiv
  e-prints, arXiv:2301.07116, \dodoi{10.48550/arXiv.2301.07116}

\bibitem[{{Springel}(2000)}]{2000MNRAS.312..859S}
{Springel}, V. 2000, \mnras, 312, 859, \dodoi{10.1046/j.1365-8711.2000.03187.x}

\bibitem[{{Springel} \& {White}(1999)}]{1999MNRAS.307..162S}
{Springel}, V., \& {White}, S. D.~M. 1999, \mnras, 307, 162,
  \dodoi{10.1046/j.1365-8711.1999.02613.x}

\bibitem[{{Stern} {et~al.}(2019){Stern}, {Fielding}, {Faucher-Gigu{\`e}re}, \&
  {Quataert}}]{2019MNRAS.488.2549S}
{Stern}, J., {Fielding}, D., {Faucher-Gigu{\`e}re}, C.-A., \& {Quataert}, E.
  2019, \mnras, 488, 2549, \dodoi{10.1093/mnras/stz1859}

\bibitem[{{Stern} {et~al.}(2023){Stern}, {Fielding}, {Hafen}, {Su}, {Naor},
  {Faucher-Gigu{\`e}re}, {Quataert}, \& {Bullock}}]{2023arXiv230600092S}
{Stern}, J., {Fielding}, D., {Hafen}, Z., {et~al.} 2023, arXiv e-prints,
  arXiv:2306.00092, \dodoi{10.48550/arXiv.2306.00092}

\bibitem[{Su(2023)}]{AGN_Jet_halo_mass}
Su, K.-Y. 2023, 14

\bibitem[{{Su} {et~al.}(2018){Su}, {Hayward}, {Hopkins}, {Quataert},
  {Faucher-Gigu{\`e}re}, \& {Kere{\v{s}}}}]{2018MNRAS.473L.111S}
{Su}, K.-Y., {Hayward}, C.~C., {Hopkins}, P.~F., {et~al.} 2018, \mnras, 473,
  L111, \dodoi{10.1093/mnrasl/slx172}

\bibitem[{{Su} {et~al.}(2017){Su}, {Hopkins}, {Hayward}, {Faucher-Gigu{\`e}re},
  {Kere{\v{s}}}, {Ma}, \& {Robles}}]{2017MNRAS.471..144S}
{Su}, K.-Y., {Hopkins}, P.~F., {Hayward}, C.~C., {et~al.} 2017, \mnras, 471,
  144, \dodoi{10.1093/mnras/stx1463}

\bibitem[{{Su} {et~al.}(2019){Su}, {Hopkins}, {Hayward}, {Ma},
  {Faucher-Gigu{\`e}re}, {Kere{\v{s}}}, {Orr}, {Chan}, \&
  {Robles}}]{2019MNRAS.487.4393S}
---. 2019, \mnras, 487, 4393, \dodoi{10.1093/mnras/stz1494}

\bibitem[{{Su} {et~al.}(2021){Su}, {Hopkins}, {Bryan}, {Somerville}, {Hayward},
  {Angl{\'e}s-Alc{\'a}zar}, {Faucher-Gigu{\`e}re}, {Wellons}, {Stern},
  {Terrazas}, {Chan}, {Orr}, {Hummels}, {Feldmann}, \&
  {Kere{\v{s}}}}]{2021MNRAS.507..175S}
{Su}, K.-Y., {Hopkins}, P.~F., {Bryan}, G.~L., {et~al.} 2021, \mnras, 507, 175,
  \dodoi{10.1093/mnras/stab2021}

\bibitem[{{Su} {et~al.}(2023){Su}, {Bryan}, {Hayward}, {Somerville}, {Hopkins},
  {Emami}, {Faucher-Gigu{\`e}re}, {Quataert}, {Ponnada}, {Fielding}, \&
  {Kere{\v{s}}}}]{2023arXiv231017692S}
{Su}, K.-Y., {Bryan}, G.~L., {Hayward}, C.~C., {et~al.} 2023, arXiv e-prints,
  arXiv:2310.17692, \dodoi{10.48550/arXiv.2310.17692}

\bibitem[{{Vogelsberger} {et~al.}(2020){Vogelsberger}, {Marinacci}, {Torrey},
  \& {Puchwein}}]{2020NatRP...2...42V}
{Vogelsberger}, M., {Marinacci}, F., {Torrey}, P., \& {Puchwein}, E. 2020,
  Nature Reviews Physics, 2, 42, \dodoi{10.1038/s42254-019-0127-2}

\bibitem[{{Weinberger} \& {Hernquist}(2023)}]{2023MNRAS.519.3011W}
{Weinberger}, R., \& {Hernquist}, L. 2023, \mnras, 519, 3011,
  \dodoi{10.1093/mnras/stac3708}

\bibitem[{{Weinberger} {et~al.}(2018){Weinberger}, {Springel}, {Pakmor},
  {Nelson}, {Genel}, {Pillepich}, {Vogelsberger}, {Marinacci}, {Naiman},
  {Torrey}, \& {Hernquist}}]{2018MNRAS.479.4056W}
{Weinberger}, R., {Springel}, V., {Pakmor}, R., {et~al.} 2018, \mnras, 479,
  4056, \dodoi{10.1093/mnras/sty1733}

\bibitem[{{Werk} {et~al.}(2016){Werk}, {Prochaska}, {Cantalupo}, {Fox},
  {Oppenheimer}, {Tumlinson}, {Tripp}, {Lehner}, \&
  {McQuinn}}]{2016ApJ...833...54W}
{Werk}, J.~K., {Prochaska}, J.~X., {Cantalupo}, S., {et~al.} 2016, \apj, 833,
  54, \dodoi{10.3847/1538-4357/833/1/54}

\bibitem[{{Yang} \& {Reynolds}(2016)}]{2016ApJ...818..181Y}
{Yang}, H.-Y.~K., \& {Reynolds}, C.~S. 2016, \apj, 818, 181,
  \dodoi{10.3847/0004-637X/818/2/181}

\bibitem[{{Yoon} {et~al.}(2018){Yoon}, {Yuan}, {Gan}, {Ostriker}, {Li}, \&
  {Ciotti}}]{2018ApJ...864....6Y}
{Yoon}, D., {Yuan}, F., {Gan}, Z.-M., {et~al.} 2018, \apj, 864, 6,
  \dodoi{10.3847/1538-4357/aad37e}

\end{thebibliography}

\end{document}